\documentclass[11pt,onecolumn,dvips,draftcls]{IEEEtran}
\usepackage{psfig, graphics, amsfonts, amsmath, color, amssymb, amsxtra, times}
\definecolor{gray}{cmyk}{.2,0.2,.3,.1}
\definecolor{dred}{cmyk}{0,0.9,0.4,0.3}
\definecolor{dblue}{rgb}{0,0,0.5}
\definecolor{dgreen}{rgb}{0,0.3,0}
\definecolor{dgray}{rgb}{0.3,0.3,0}
\usepackage[breaklinks=true, colorlinks=true, linkcolor=black, urlcolor=dblue,
  citecolor=black, pdfpagemode=None, pdfstartview=FitH]{hyperref}

\DeclareOldFontCommand{\rm}{\normalfont\rmfamily}{\mathrm}
\DeclareOldFontCommand{\sf}{\normalfont\sffamily}{\mathsf}
\DeclareOldFontCommand{\tt}{\normalfont\ttfamily}{\mathtt}
\DeclareOldFontCommand{\bf}{\normalfont\bfseries}{\mathbf}
\DeclareOldFontCommand{\it}{\normalfont\itshape}{\mathit}
\DeclareOldFontCommand{\sl}{\normalfont\slshape}{\@nomath\sl}
\DeclareOldFontCommand{\sc}{\normalfont\scshape}{\@nomath\sc}

\newtheorem{theorem}{Theorem}
\newtheorem{proposition}{Proposition}
\newtheorem{lemma}{Lemma}
\newcommand{\rend}{\hfill$\square$}
\newcommand{\tend}{\hfill$\blacksquare$}

\newcommand{\expect}[1]{\ensuremath{\operatorname{E}\left[#1\right]}}

\title{Multiterminal Source Coding With Two Encoders--I: A Computable
  Outer Bound
  \thanks{The author is with the School of Electrical and Computer
  Engineering, Cornell University, Ithaca, NY.  URL:
  \href{http://cn.ece.cornell.edu/}{{\tt http://cn.ece.cornell.edu/}}.
  Work supported by the National Science Foundation, under awards
  CCR-0238271 (CAREER), CCR-0330059, and ANR-0325556.}}
\author{Sergio D.\ Servetto}
\date{November 12, 2006.}

\begin{document}
\maketitle
\thispagestyle{empty}

\begin{picture}(0,0)
\put(-5,210){\tt\small Submitted to the IEEE Transactions on Information
  Theory, April 2006;  Revised,}
\put(-5,200){\tt\small November 2006.}
\end{picture}
\vspace{-4mm}

\begin{abstract}
\noindent\it
In this first part, a computable outer bound is proved for the
multiterminal source coding problem, for a setup with two encoders,
discrete memoryless sources, and bounded distortion measures.
\end{abstract}

\vspace{1cm}

\vspace{9.3cm}
\pagebreak
\setcounter{page}{1}

\section{Introduction}

\subsection{The Problem of Multiterminal Source Coding}

\begin{figure}[!ht]
\centerline{\psfig{file=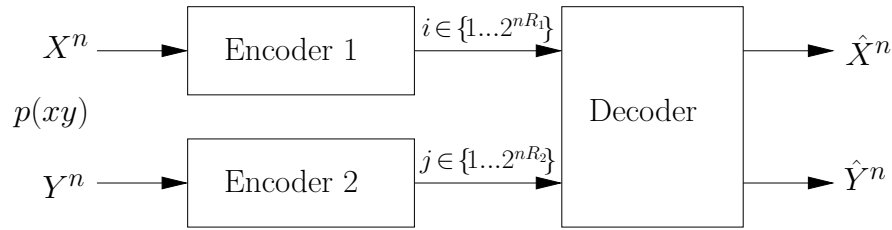,height=3cm,width=12cm}}
\caption{System setup for multiterminal source coding.}
\label{fig:setup}
\end{figure}

Consider two dependent sources $X$ and $Y$, with joint distribution
$p(xy)$.  These sources are to be encoded by two separate encoders,
each of which observes only one of them, and are to be decoded by a
single joint decoder.  $X$ is encoded at rate $R_1$ and with average
distortion $D_1$, and $Y$ is encoded at rate $R_2$ and with average
distortion $D_2$.  This setup is illustrated in Fig.~\ref{fig:setup}.

In the classical {\em multiterminal source coding} problem, as
formulated in~\cite{Berger:78, Tung:PhD}, the goal
is to determine the region of all achievable rate-distortion tuples
$(R_1,R_2,D_1,D_2)$.  Although relatively simple to describe (a 
formal description is given later), the multiterminal source coding
problem was one of the long-standing open problems in information
theory -- see, e.g.,~\cite[pg.\ 443]{CoverT:91}.  Furthermore,
besides its historical interest, this problem also comes up naturally
in the context of a sensor networking problem of interest to
us~\cite{BarrosS:06}.

Multiterminal source coding has rich history, among which
fundamental contributions, in chronological order, are the
works of: a) Dobrushin-Tsybakhov~\cite{DobrushinT:62}, with the
first rate-distortion problem with a Markov chain constraint; b)
Slepian-Wolf~\cite{SlepianW:73b}, with the formulation and solution
to the first distributed source coding problem, and
Cover~\cite{Cover:75b}, with a simpler proof of the Slepian-Wolf
result, a proof method widely in use today; c)
Ahlswede-K\"orner~\cite{AhlswedeK:75} and Wyner~\cite{Wyner:75},
with the first use of an auxiliary random variable to describe
the rate region of a source coding problem, and with it the need
to introduce proof methods to bound their cardinality; d)
Wyner-Ziv~\cite{WynerZ:76}, with the first characterization of a
multiterminal rate-distortion function; e) Berger-Tung~\cite{Berger:78,
Tung:PhD}, with the first formulation and partial results on the
multiterminal source coding problem as formulated in Fig.~\ref{fig:setup};
and f) Berger-Yeung~\cite{BergerY:89, Yeung:PhD}, with a complete
solution to a more general form of the Wyner-Ziv problem.  For
details on these, and on {\em many} more important contributions,
as well as for historical information on the problem, the reader
is referred to~\cite{BergerS:07}.  

\begin{figure}[ht]
\centerline{\resizebox{10cm}{6cm}{\input{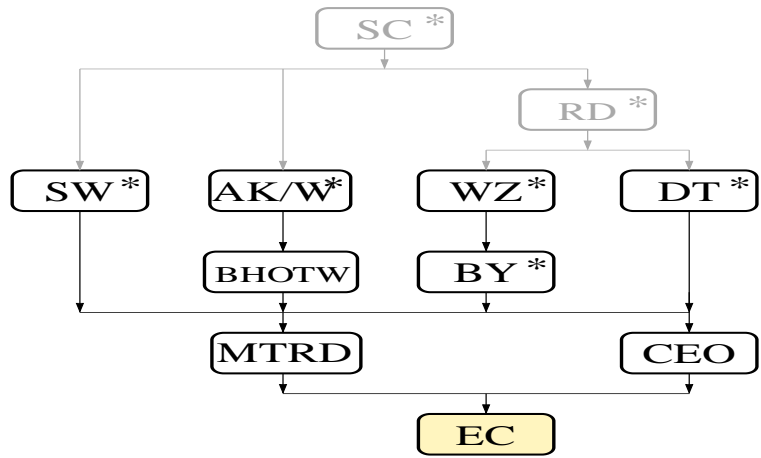}}}
\caption{A hierarchy of problems in multiterminal source coding
  with two encoders and one decoder: an arrow from problem X to
  problem Y indicates that X is a special case of Y, in the sense
  that a solution to Y automatically provides a solution to X.
  Abbreviations -- SC: two-terminal lossless source coding;
  RD: two-terminal rate-distortion~\cite{Shannon:59}; SW: distributed
  coding of dependent sources~\cite{SlepianW:73b}; AK/W: source
  coding with side information~\cite{AhlswedeK:75, Wyner:75};
  WZ: rate-distortion with side information~\cite{WynerZ:76};
  BY: the Berger-Yeung extension of WZ theory~\cite{BergerY:89};
  DT: rate-distortion with a remote source~\cite{DobrushinT:62};
  BHOTW: a rate-distortion formulation of the Ahlswede-K\"orner-Wyner
  problem~\cite{BergerHOTW:79}; CEO: the CEO problem~\cite{BergerZV:96};
  MTRD: the problem setup of Fig.~\ref{fig:setup}; EC: the entropy
  characterization problem~\cite{CsiszarK:80}.  Asterisks are used
  to indicate problems whose solution was previously known.}
\label{fig:hierarchy}
\end{figure}

The setup of Fig.~\ref{fig:setup} represents what we feel was the
simplest yet unsolved instance of a multiterminal source coding problem.
The problem of Fig.~\ref{fig:setup}, and the CEO problem~\cite{BergerZV:96}
are, to the best of our knowledge, the last two known special cases of
the general entropy characterization of problem of Csisz\'ar and
K\"orner~\cite{CsiszarK:80} that remained unsolved.  This hierarchy
of problems is illustrated in Fig.~\ref{fig:hierarchy}.

It should be pointed out though that the setup of Fig.~\ref{fig:setup}
is by no means the most general formulation of a multiterminal source
coding problem we could have given, there are many other ways in which
we could have chosen to formulate these problems: we could have chosen
a network with $M$ encoders and a single decoder which attempts to
reconstruct $L$ different functions of the sources, we could have
considered continuous-alphabet and/or general ergodic sources, we
could have considered feedback and interactive communication, we could
have studied how this problem relates to the network coding problem,
and we could have considered network
topologies with multiple decoders as well.  All these alternative
possible formulations are discussed in detail in~\cite{BergerS:07}.

\subsection{Difficulties in Proving a Converse}
\label{sec:difficulties}

Among the limited number of references mentioned above, we included
the Berger-Tung bounds~\cite{Berger:78, Tung:PhD}.  These bounds do
provide the best known descriptions of the region of achievable rates
for the problem setup of Fig.~\ref{fig:setup},\footnote{We note that
recently, a new outer bound has been proposed for a version of
multiterminal source coding
that contains the formulation of~\cite{Berger:78, Tung:PhD} considered
here as a special case~\cite{Wagner:PhD, WagnerA:05}.  The new
bound has many desirable properties: it unifies known bounds custom
developed for seemingly different problems, and it provides a conclusive
answer for a previously unsolved instance.  However, when specialized
to our two-encoder setup, it is unclear if the new bound provides
an improvement over the Berger-Tung outer bound.  So, due to the
simplicity of the latter, we have chosen here to focus on that one
instead of on the more modern form.}
and so we elaborate on those now.

\medskip\begin{proposition}[Berger-Tung Bounds]
\label{prp:bt-bounds}
Fix $(D_1,D_2)$.  Let $X$ and $Y$ be two sources out of which
pairs of sequences $\big(X^n,Y^n\big)$ are drawn i.i.d.~$\sim p(xy)$;
and let $U$ and $V$ be auxiliary variables defined over alphabets
$\mathcal{U}$ and $\mathcal{V}$, such that there exist functions
$\gamma_1:\mathcal{U}\times\mathcal{V}\to\hat{\mathcal{X}}$
and $\gamma_2:\mathcal{U}\times\mathcal{V}\to\hat{\mathcal{Y}}$,
for which $\expect{d_1\big(X,\gamma_1(UV)\big)}\leq D_1$ and
$\expect{d_2\big(Y,\gamma_2(UV)\big)}\leq D_2$.  Consider rates
$(R_1,R_2)$, such that $R_1 \geq I(XY\wedge U|V)$,
$R_2 \geq I(XY\wedge V|U)$, and $R_1+R_2 \geq I(XY\wedge UV)$,
for some joint distribution $p(xyuv)$.  Now:
\begin{itemize}
\item for any $p(xyuv)$ that satisfies a Markov chain of the form
  $U-X-Y-V$, all rates $(R_1,R_2)$ obtained for any such
  $p$ are achievable;
\item if there exists a $p(xyuv)$ that satisfies two Markov chains of
  the form $U-X-Y$ and $X-Y-V$, then if we consider the union of the
  set of rates defined for each such $p(xyuv)$, we must have that any
  achievable rates are included in that union;
\end{itemize}
that is, the first condition defines an {\em inner} bound, and the second
an {\em outer} bound to the rate region. \rend
\end{proposition}\medskip

The regions defined by these bounds, when regarded as images of maps
that transform probability distributions into rate pairs, have a
property that is a source of many difficulties: the mutual information
expressions that define the inner and the outer bounds are identical,
it is only the {\em domains} of the two maps that differ; as such,
comparing the resulting regions is difficult.  This difference between
the inner and outer bounds has been the state of affairs in multiterminal
source coding, since 1978.

A close examination of these distributions suggested to us that the
gap might not be due to a suboptimal coding strategy used in the inner
bound, but instead that perhaps the outer bound allows for the inclusion
of dependencies that cannot be physically realized by any distributed
code.  Consider these distributions:
\begin{itemize}
\item For the inner bound, $p(xyuv)$ = $p(xy)p(u|x)p(v|y)$.
\item For the outer bound, $p(xyuv)$
  = $p(xy)p(u|x)p(v|y\underline{xu})$
  = $p(xy)p(v|y)p(u|x\underline{yv})$.
\end{itemize}
If we choose to interpret $U$ and $V$ as instantaneous descriptions
of encodings of $X$ and $Y$,
then we see that the outer bound says that the encoding
$V$ is allowed to contain information about $X$ {\em beyond} that
which can be extracted from $Y$, and likewise for $U$ and
$Y$.\footnote{Note: this interpretation comes from the inner bound,
and is only justified for {\em blocks}.  $U^n$ does represent an encoding
of $X^n$, but it would be incorrect to say that the variable $U$ is
an encoding of $X$ (and likewise for $V$ and $Y$).  These insights
can only be carried so far, but at this point we are only trying to
build some intuition, and thus it is permissible to take such liberties.}
Motivated by this observation, in the first part of this work we
set ourselves the goal of finding a new outer bound.

\subsection{An Interpretation of Distributed Rate-Distortion Codes
  as Constrained Source Covers}

In Part I of this paper we present a finitely parameterized outer
bound for the region of achievable rates of the multiterminal source
coding problem of Fig.~\ref{fig:setup}, based on what we
believe is an original proof technique.  Some highlights of that
proof method, formally developed in later sections, are provided here.

\subsubsection{Rate-Distortion Codes $\equiv$ Source Covers}
\label{sec:intro-distributed-covers}

Our proof tightens existing converses by means of identifying a
constraint that {\em all} codes are subject to, but that is not
captured by any existing outer bound.  To explain what the constraint
is, the easiest way to get started is by drawing an analogy to
classical, two-terminal rate-distortion codes.

\begin{figure}[ht]
\centerline{\resizebox{15cm}{4cm}{\input{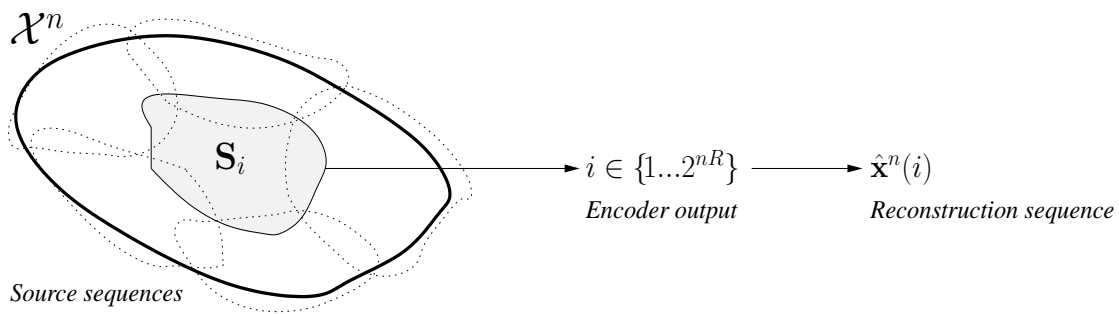}}}
\vspace{-2mm}
\caption{Cover-based representation of a classical rate-distortion code.}
\label{fig:covers-classical}
\end{figure}

In the standard, two-terminal rate-distortion problem, a generic
code consists of the following elements:
\begin{itemize}
\item A block length $n$.
\item A cover $\big\{ \mathbf{S}_i \;:\; i=1...2^{nR} \big\}$ of the source
  $\mathcal{X}^n$.
\item A reconstruction sequence $\hat{\mathbf{x}}^n(i)$, associated to each
  cover element $\mathbf{S}_i$.
\end{itemize}
Given this description, an encoder $f:\mathcal{X}^n\to\{1...2^{nR}\}$
makes $f\big(\mathbf{x}^n\big)=i$ for some source sequence $\mathbf{x}^n$
and some index $i$, if
$\mathbf{x}^n\in \mathbf{S}_i$, with ties broken arbitrarily; a decoder
$g:\{1...2^{nR}\}\to
\hat{\mathcal{X}}^n$ simply maps $g(i)=\hat{\mathbf{x}}^n(i)$.  And we say
that the encoder/decoder pair $(f,g)$ satisfies a distortion constraint $D$
if, roughly, $P\Big(d\big(\mathbf{x}^n,g(f(\mathbf{x}^n))\big)\leq D\Big)
\approx 1$, for all $n$ large enough.  Such a representation is illustrated
in Fig.~\ref{fig:covers-classical}.

\begin{figure}[ht]
\centerline{\psfig{file=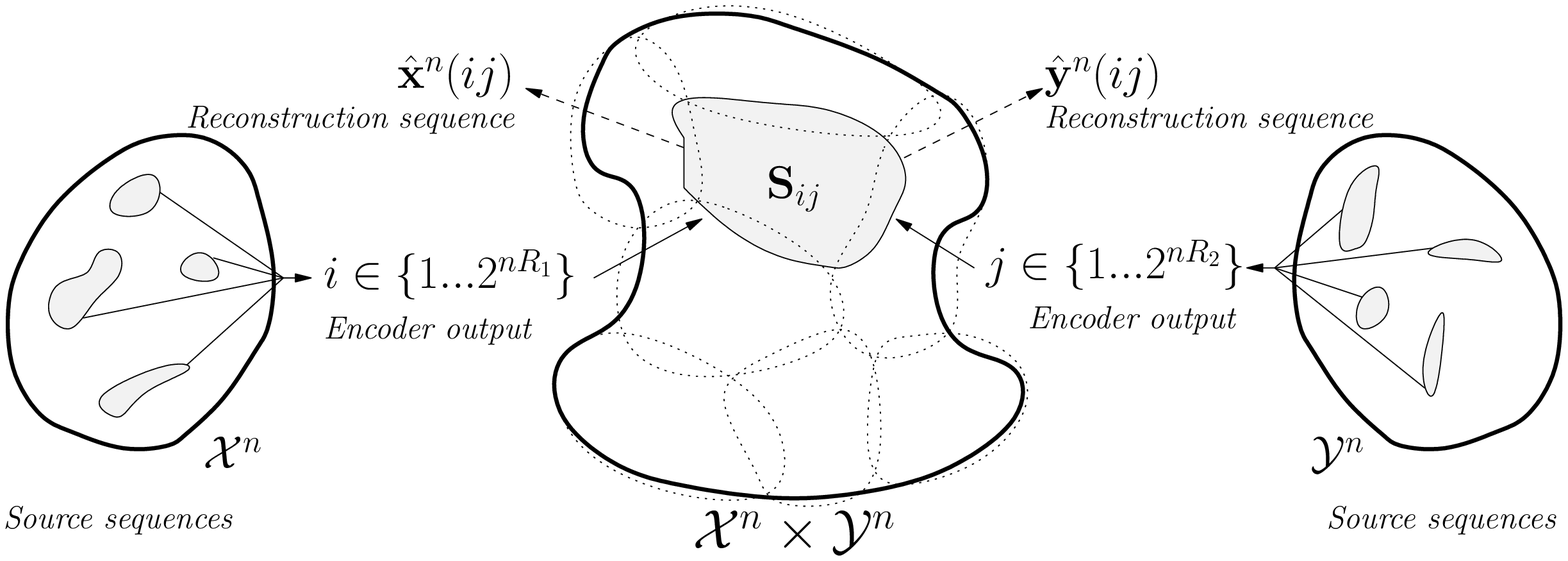,height=6cm,width=16cm}}
\caption{Cover-based representation of a {\em distributed} rate-distortion
  code.}
\label{fig:covers-distributed}
\end{figure}

In an analogous manner, we specify an arbitrary {\em distributed}
rate-distortion code as follows:
\begin{itemize}
\item A block length $n$.
\item {\em Two} covers:
  \begin{itemize}
  \item A cover $\big\{ \mathbf{S}_{1,i} \;:\; i=1...2^{nR_1}\big\}$ of the
    source $\mathcal{X}^n$.
  \item A cover $\big\{ \mathbf{S}_{2,j} \;:\; j=1...2^{nR_2}\big\}$ of the
    source $\mathcal{Y}^n$.
  \end{itemize}
  Indirectly, these two covers specify a cover $\mathbf{S}_{ij}\;\triangleq\;
    \big\{ \mathbf{S}_{1,i}\times\mathbf{S}_{2,j} : i=1...2^{nR_1},
    j=1...2^{nR_2}\big\}$ of the product alphabet $\mathcal{X}^n\times
    \mathcal{Y}^n$.
\item For each cover element $\mathbf{S}_{ij}$, we specify
  {\em two} reconstruction sequences
  $\big(\hat{\mathbf{x}}^n(ij),\hat{\mathbf{y}}^n(ij)\big)$.
\end{itemize}
Given this description, an encoder $f_1:\mathcal{X}^n\to\{1...2^{nR_1}\}$
for node 1 makes $f_1\big(\mathbf{x}^n\big)=i$ for some source sequence
$\mathbf{x}^n$ and some index $i$, if $\mathbf{x}^n\in\mathbf{S}_{1,i}$,
with ties broken arbitrarily (and similarly for an encoder $f_2$ at node 2);
a decoder $g:\{1...2^{nR_1}\}\times\{1...2^{nR_2}\}\to
\hat{\mathcal{X}}^n\times\hat{\mathcal{Y}}^n$ simply maps
$g(i,j)=\big(\hat{\mathbf{x}}^n(ij),\hat{\mathbf{y}}^n(ij)\big)$.
And we say that the distributed code $(f_1,f_2,g)$ satisfies two distortion
constraints $D_1$ and $D_2$ if, roughly,
$P\Big(d_1\big(\mathbf{x}^n,\hat{\mathbf{x}}^n\big)\leq D_1
\mbox{ and }
d_2\big(\mathbf{y}^n,\hat{\mathbf{y}}^n\big)\leq D_2\Big)\approx 1$,
for all $n$ large enough, and for
$\big(\hat{\mathbf{x}}^n\hat{\mathbf{y}}^n\big)=
g\big(f_1(\mathbf{x}^n),f_2(\mathbf{y}^n)\big)$.  Such a representation is
illustrated in Fig.~\ref{fig:covers-distributed}.

\subsubsection{Constraints on the Structure of Source Covers}

\begin{figure}[!h]
\centerline{\psfig{file=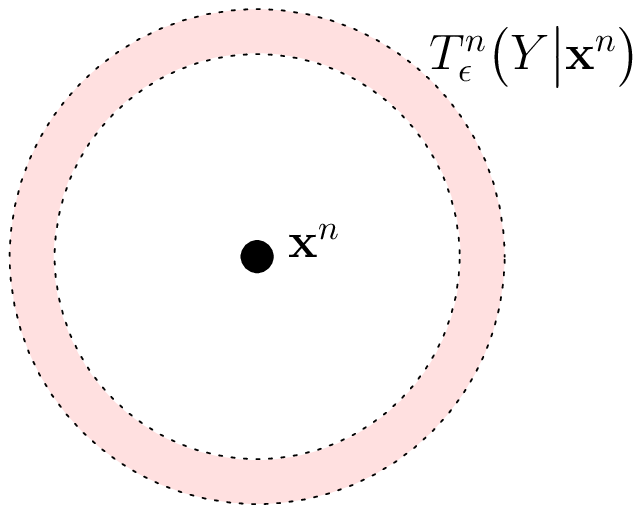,height=6cm}\hspace{1cm}
            \psfig{file=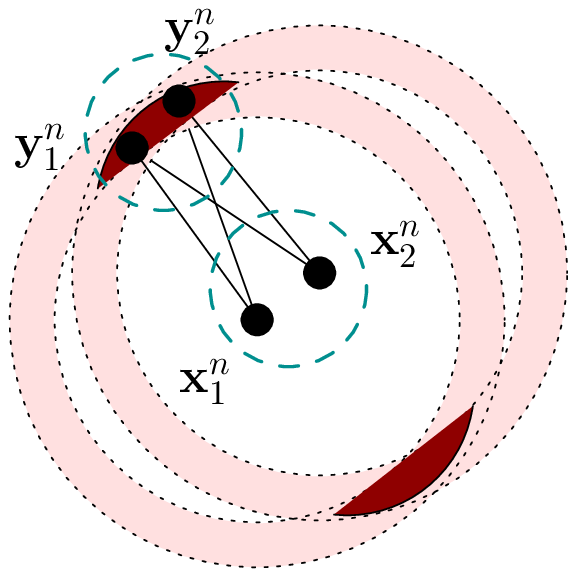,height=6cm}}
\caption{An example, to illustrate the significance of the
  requirement that cover elements $\mathbf{S}_{ij}$ take a product form.
  Let $\mathcal{X}=\mathcal{Y}=\{0,1\}$, and $p(xy)=p(x)p(y|x)$
  specified by a $p(x)$ such that $P(X=0)=P(X=1)=\frac 1 2$,
  and $p(y|x)$ a binary symmetric channel with crossover probability
  $p_c$.  Left: for each typical $\mathbf{x}^n$, there is a ``ring'' of
  $\mathbf{y}^n$'s jointly typical with it, centered at $\mathbf{x}^n$
  and of radius $\approx np_c$.  Right: consider pairs
  $\big(\mathbf{x}_1^n\mathbf{y}_1^n\big)$ and $\big(\mathbf{x}_2^n
  \mathbf{y}_2^n\big)$ in $\mathbf{S}_{ij}$; dashed circles denote
  distortion balls centered at $\hat{\mathbf{x}}^n(ij)$ and
  $\hat{\mathbf{y}}^n(ij)$ (with the centers omitted, for clarity),
  and dark shaded regions denote the intersection of two rings.
  Suppose now that all four pairs $(\mathbf{x}_1^n\mathbf{y}_1^n)$,
  $(\mathbf{x}_1^n\mathbf{y}_2^n)$ $(\mathbf{x}_2^n\mathbf{y}_1^n)$,
  and $(\mathbf{x}_2^n\mathbf{y}_2^n)$ are in $T_\epsilon^n\big(XY\big)$.
  Because $\mathbf{S}_{ij}= \mathbf{S}_{1,i}\times\mathbf{S}_{2,j}$,
  {\em all four pairs must be in $\mathbf{S}_{ij}$ as well:} the decoder
  does not have enough information to discriminate among these pairs.
  No such constraint exists with a centralized encoder.}
\label{fig:binary-example}
\end{figure}

Our main insight is that, whereas in the classical problem
any arbitrary cover defines a valid rate-distortion code, in multiterminal
source coding this is no longer the case: {\em covers of the product source
$\mathcal{X}^n\times\mathcal{Y}^n$ only of the form $\mathbf{S}_{ij}
= \mathbf{S}_{1,i}\times\mathbf{S}_{2,j}$ can be realized by distributed
codes}.  The significance of this requirement is illustrated with an
example in Fig.~\ref{fig:binary-example}.

From the informal argument of Fig.~\ref{fig:binary-example},
we see how the fact that distributed codes produce covers only of the
form $\mathbf{S}_{ij}=\mathbf{S}_{1,i}\times\mathbf{S}_{2,j}$ results
in constraints on the sets used to cover the typical set
$T_\epsilon^n\big(XY\big)$: there are certain groups of typical
sequences that cannot be broken, in the sense that either all of them
appear together in a cover element $\mathbf{S}_{ij}$, or none of them
appear.  We believe this is significant for two main reasons:
\begin{itemize}
\item If we compare to a classical rate-distortion code, this constraint
  is clearly not there.  Provided the distortion constraints are met, a
  classical code would be able to split the typical set into distortion
  balls, without any further constraints.
\item More fundamentally though, we view this constraint as a form of
  ``independence,'' reminiscent to us of the extra independence assumption
  required by the long Markov chain used in the definition of the
  Berger-Tung inner bound, which is not there in the definition of the
  outer bound, as highlighted in Section~\ref{sec:difficulties} earlier.
\end{itemize}
This latter observation is perhaps the strongest piece of evidence that
suggested to us that the Berger-Tung inner bound might be tight.

\subsection{Main Contributions and Organization of the Paper}

The main contribution presented in Part I of this paper is the
development of an outer bound to the region of achievable rates
for multiterminal source coding.  This outer bound has two salient
properties that distinguish it from existing bounds in the literature:
\begin{itemize}
\item it is based on explicitly modeling a constraint on the
  structure of codes that, as we understand things, had not been
  captured by any previously developed bound;
\item and also unlike existing bounds, it is finitely parameterized.
\end{itemize}
We believe that this outer bound coincides with the set of achievable
rates defined by the Berger-Tung inner bound.  This issue is thoroughly
explored in Part II of this paper, in the context of our study of
algorithmic issues involved in the effective computation of this bound.

The rest of this paper is organized as follows.  In
Section~\ref{sec:preliminaries} we define our notation, and state
our main result.  In Section~\ref{sec:aux-lemmas} we state and prove
some auxiliary lemmas that greatly simplify the proof of the main
theorem, a proof that is fully developed in Section~\ref{sec:main-proof}.
The paper concludes with an extensive discussion on our main result
and its implications, in Section~\ref{sec:discussion}.

\section{Preliminaries}
\label{sec:preliminaries}

\subsection{Definitions and Notation}

First, a word about notation.  Random variables are denoted with
capital letters, e.g., $X$.  Realizations of these variables are
denoted with lower case letters: e.g., $X=x$ means that the random
variable $X$ takes on the value $x$.  Script letters are typically
used to denote alphabets, e.g., the random variable $X$ takes values
on an alphabet $\mathcal{X}$.  The alphabets of all random variables
considered in this work are always assumed finite.  Sets in general
are denoted by capital boldface symbols, e.g., $\mathbf{S}$.
The size of a set is denoted by $\big|\mathbf{S}\big|$.  A
probability mass function on $\mathcal{X}$ is denoted by $p_X(x)$,
or simply $p(x)$ when the variable that it applies to is clear from
the context.  Sequences of elements from an alphabet $\mathcal{X}$
are denoted by boldface symbols $\mathbf{x}^n$,
and its $i$-th element by $\mathbf{x}_i$; this sequence is an element
of the extension alphabet $\mathcal{X}^n$.  The expression
$\mathbf{x}_i^{j,n}$ denotes a subsequence of $\mathbf{x}^n$ consisting
of the elements $[\mathbf{x}_i,\mathbf{x}_{i+1},...,\mathbf{x}_j]$,
whenever $i\leq j$, otherwise it denotes an empty sequence; also,
sometimes the length $n$ of the sequence will be clear from the
context, and then we simply write $\mathbf{x}_i^j$ instead of
$\mathbf{x}_i^{j,n}$, whenever this does not cause confusion.  The
expression $\mathbf{x}^{-i,n}$ denotes the sequence
$[\mathbf{x}_1,...,\mathbf{x}_{i-1},\mathbf{x}_{i+1},...,\mathbf{x}_n]$,
and again, we write this as $\mathbf{x}^{-i}$ whenever $n$ is
clear from the context.  The same conventions are followed for
sequences of random variables.

Given a boolean predicate $b(\mathbf{x})$ depending on a variable
$\mathbf{x}$, we write $1_{\{b(\mathbf{x})\}}$ to denote
the indicator function for the predicate: this is a function that
takes the value 1 whenever $b(\mathbf{x})$ is true, and 0 whenever
it is false.  Given a sequence $\mathbf{x}^n\in\mathcal{X}^n$,
and an element $x\in\mathcal{X}$, we denote by $N(x;\mathbf{x}^n)$
the type of $\mathbf{x}^n$, defined as
$N(x;\mathbf{x}^n)=\sum_{i=1}^n 1_{\{\mathbf{x}_i=x\}}$.  Then,
for any random variable $X$, any real number $\epsilon>0$, and
any integer $n>0$, we denote by $T_\epsilon^n(X)$ the strongly typical
set of $X$ with parameters $n$ and $\epsilon$, defined as
\[ T_\epsilon^n(X) \;\;=\;\; \Big\{ \mathbf{x}^n\in\mathcal{X}^n \;\Big|\;
   \forall x\in\mathcal{X}:
   \big|\mbox{$\frac 1 n$}N(x;\mathbf{x}^n)-p_X(x)\big|
   < \mbox{$\frac\epsilon{|\mathcal{X}|}$} \Big\}.
\]
In some situations, we need to compare typical sets defined for
the same set of variables, but induced by different distributions on
these variables.  To resolve this ambiguity, we denote by
$T_\epsilon^n\big(X\big)[p_X]$ the typical set corresponding to a
distribution $p_X$.  The same convention is followed when there is
similar ambiguity in the evaluation of entropies (denoted
$H\big(X\big)[p_X]$), and mutual information expressions (denoted
$I\big(X\wedge Y\big)[p_{XY}]$).

Vector extensions $N(xy;\mathbf{x}^n\mathbf{y}^n)$, $T_\epsilon^n(XY)$,
etc., are defined by considering the same definitions as above, over a
suitable product alphablet $\mathcal{X}\times\mathcal{Y}$.  Similarly,
given two random variables $X$ and $Y$, a joint probability mass
function $p_{XY}(xy)$,
and a sequence $\mathbf{y}^n$, we denote by $T_\epsilon^n(X|\mathbf{y}^n)$
the conditional typical set of $X$ given $\mathbf{y}^n$, defined as
\[ T_\epsilon^n\big(X\big|\mathbf{y}^n\big)
   \;\;=\;\; \Big\{ \mathbf{x}^n\in\mathcal{X}^n \;\Big|\;
   \forall x\in\mathcal{X},y\in\mathcal{Y}:
   \big|\mbox{$\frac 1 n$}N(xy;\mathbf{x}^n\mathbf{y}^n)-p_{XY}(xy)\big|
   < \mbox{$\frac\epsilon{|\mathcal{X}||\mathcal{Y}|}$} \Big\}.
\]
We will also consider situations where we need to refer to the set of
all typical sequences which are jointly typical with at least one of a
group.  In that case, for a set $\mathbf{S}\subseteq\mathcal{Y}^n$, we
write
\[ T_\epsilon^n\big(X\big|\mathbf{S}\big)
   \;\;=\;\; \bigcup_{\mathbf{y}^n\in\mathbf{S}}
             T_\epsilon^n\big(X\big|\mathbf{y}^n\big).
\]

Given any $\epsilon>0$, many times we require to make reference
to quantities which are deterministic functions of $\epsilon$, having
the property that as $\epsilon\to 0$, these quantities also vanish.
Such small quantities are denoted by $\epsilon_1$, $\epsilon_2$,
$\dot\epsilon$, $\ddot\epsilon$, $\epsilon'$, $\epsilon''$, etc.;
and the value of $\epsilon$ on which they depend is either mentioned
explicitly or should be clear from the context.

Consider two random variables
$X$ and $Y$ with joint distribution $p(xy)$.  $T_\epsilon^n\big(X)$
is the usual typical set.  Sometimes we also need to consider
the set $S_{\epsilon,Y}^n(X)\triangleq\Big\{\mathbf{x}^n\,\Big|\,
T_\epsilon^n\big(Y\big|\mathbf{x}^n\big)\neq\emptyset\Big\}$.  Clearly,
$S_{\epsilon,Y}^n(X)\subseteq T_\epsilon^n\big(X)$.  But we also
know from~\cite[Ch.\ 5]{Yeung:01}, that
$\Big|\frac 1 n\log\big|S_{\epsilon,Y}^n(X)\big|-H(X)\Big|<\dot\epsilon$.
That is, although there may exist strongly typical sequences $\mathbf{x}^n$
for which there are no sequences $\mathbf{y}^n$ jointly typical with them,
these $\mathbf{x}^n$'s form a set of vanishing measure.

Some standard operations on sets are intersection
($\mathbf{A}\cap\mathbf{B}$), union ($\mathbf{A}\cup\mathbf{B}$),
complementation ($\mathbf{A}^c$) and difference
($\mathbf{A}\backslash\mathbf{B}$).   The set of all subsets of
$\mathbf{S}$ is denoted by $2^{\mathbf{S}}$.  The convex closure of $\mathbf{S}$
is denoted by
$\overline{\mathbf{S}}=\bigcap\big\{\mathbf{S}'\;\big|\;\mathbf{S}\subseteq
\mathbf{S}'\,\wedge\,\mathbf{S}'\mbox{ is closed and convex}\big\}$.
Given a set $\mathbf{S}$, a cover of size $N$ of $\mathbf{S}$ is a
collection of sets $\mathcal{S}=\big\{\mathbf{S}_i:i=1...N\big\}$,
such that $\mathbf{S}\subseteq\bigcup_{i=1}^N\mathbf{S}_i$.  If a
cover further satisfies that $\mathbf{S}_i\cap \mathbf{S}_j=\emptyset$
($1\leq i\neq j\leq N$), and that $\mathbf{S}=\bigcup_{i=1}^N
\mathbf{S}_i$, then we say that $\mathcal{S}$ is a {\em partition}
of $\mathbf{S}$.

Consider two sets, $\mathbf{A}$ and $\mathbf{B}$, for which
$P\big(\mathbf{B}\big|\mathbf{A}\big)=1$: clearly,
$P\big(\mathbf{A}\cap\mathbf{B}\big)=P\big(\mathbf{A}\big)$,
and hence $\mathbf{A}\subseteq\mathbf{B}$, except perhaps for
a set of measure zero.  If instead we have a slightly weaker
condition, namely that $P\big(\mathbf{B}\big|\mathbf{A}\big)>1-\epsilon$,
then we say that $\mathbf{A}$ is {\em weakly included} in $\mathbf{B}$,
and we denote this by $\mathbf{A}\subseteq_\epsilon\mathbf{B}$.

\subsection{Distributed Rate-Distortion Codes}

Consider two sources $X$ and $Y$, out of which random pairs of
sequences $\big(X^n,Y^n\big)$ are drawn i.i.d.~$\sim p(xy)$ from two
finite alphabets, denoted $\mathcal{X}$ and
$\mathcal{Y}$, and reproduced with elements of two other alphabets
$\hat{\mathcal{X}}$ and $\hat{\mathcal{Y}}$.  The two sources
$X$ and $Y$ are processed by two separate encoders.  The
{\em encoders} are two functions:
\[ f_1:\; \mathcal{X}^n \;\;\to\;\; \big\{1,2,\dots,2^{nR_1}\big\}
   \mbox{\hspace{1cm}and\hspace{1cm}}
   f_2:\; \mathcal{Y}^n \;\;\to\;\; \big\{1,2,\dots,2^{nR_2}\big\}.
\]
These encoding functions map a block of $n$ source symbols to discrete
indices.  The {\em decoder} is a function
\[ g:\;\big\{1,2,\dots,2^{nR_1}\big\}\times\big\{1,2,\dots,2^{nR_2}\big\}
       \;\;\to\;\; \hat{\mathcal{X}}^n \times \hat{\mathcal{Y}}^n, \]
which maps a pair of indices into two blocks of reconstructed
source sequences.

Two distortion
measures $d_1:\mathcal{X}\times\hat{\mathcal{X}}\to[0,\infty)$ and
$d_2:\mathcal{Y}\times\hat{\mathcal{Y}}\to[0,\infty)$ are used to
define reconstruction quality.  Since $\infty$ is not in their
range and the alphabets are finite, these distortion measures
are necessarily bounded, so we denote these largest values by
$\max\limits_{x\in\mathcal{X},\hat x\in\hat{\mathcal{X}}}
d_1(x,\hat x)\triangleq d_{1,\mbox{\tiny MAX}}$,
$\max\limits_{y\in\mathcal{Y},\hat y\in\hat{\mathcal{Y}}}
d_2(y,\hat y) \triangleq d_{2,\mbox{\tiny MAX}}$, and
$\max\big(d_{1,\mbox{\tiny MAX}},d_{2,\mbox{\tiny MAX}}\big)
\triangleq d_{\mbox{\tiny MAX}}<\infty$.
$d_1^n\big(\mathbf{x}^n,\hat{\mathbf{x}}^n\big)
\triangleq\frac 1 n\sum_{i=1}^n d_1\big(x_i,\hat x_i\big)$
and $d_2^n\big(\mathbf{y}^n,\hat{\mathbf{y}}^n\big)
\triangleq\frac 1 n\sum_{i=1}^n d_2\big(y_i,\hat y_i\big)$
denote the corresponding extensions to blocks.  Oftentimes, the
symbols $d_1$ and $d_2$ are used for both the single-letter
and the block extensions; which is the intended meaning should
be clear from the context.  For any distortion measure
$d:\mathcal{X}^n\times\hat{\mathcal{X}}^n\to[0,\infty)$, an element
$\hat{\mathbf{x}}^n\in\hat{\mathcal{X}}^n$ and a number $D\geq 0$,
a ``ball'' of radius $D$ centered at $\hat{\mathbf{x}}^n$ is the
set $B\big(\hat{\mathbf{x}}^n,D\big)=\big\{\mathbf{x}^n\in\mathcal{X}^n
\,\big|\,d\big(\mathbf{x}^n,\hat{\mathbf{x}}^n\big))<D\big\}$
(and similarly for a ball $B\big(\hat{\mathbf{y}}^n,D\big)$).
For any $D$, $D^+$ is shorthand for $D+\dot\epsilon$, for an
$\epsilon$ that is always clear from the context.

Fix now encoders and decoder $(f_1,f_2,g)$ operating on blocks of length
$n$, and a real number $\epsilon>0$.  If we have that
\begin{equation}
   P\Big( \Big\{ \big(\mathbf{x}^n\mathbf{y}^n\big) \;\Big|\;
          \big(\hat{\mathbf{x}}^n\hat{\mathbf{y}}^n\big)
          = g\big(f_1(\mathbf{x}^n),f_2(\mathbf{y}^n)\big) \,\wedge\,
          d_1\big(\mathbf{x}^n,\hat{\mathbf{x}}^n\big)
          < D_1^+ \,\wedge\,
          d_2\big(\mathbf{y}^n,\hat{\mathbf{y}}^n\big)
          < D_2^+ \Big\}
    \Big) \;\;\geq\;\; 1-\dot{\epsilon},
  \label{eq:distortion-constraint}
\end{equation}
then we say that $(f_1,f_2,g)$ satisfies the $(\epsilon,D_1,D_2)$-distortion
constraint.\footnote{This form of a distortion constraint is referred
to as an {\it $\epsilon$-fidelity criterion} in~\cite[pg.\ 123]{CsiszarK:81}.
An alternative form to this ``local'' condition is given by requiring a
``global'' average constraint of the form
$\expect{d_1\big(\mathbf{x}^n,\hat{\mathbf{x}}^n\big)}<D_1^+$ and
$\expect{d_2\big(\mathbf{y}^n,\hat{\mathbf{y}}^n\big)}<D_2^+$.  For
the purpose of our developments, the local form lends itself more
readily to analysis, and hence is the one we adopt.}

\subsection{Achievable Rates}

A $\big(2^{nR_1},2^{nR_2},n,\epsilon,D_1,D_2\big)$ distributed
rate-distortion code is defined by a block length $n$, a
parameter $\epsilon>0$, two encoding functions $f_1$ and $f_2$
with ranges of size $2^{nR_1}$ and $2^{nR_2}$, and a decoding
function $g$, such that $(f_1,f_2,g)$ satisfies the
$\big(\epsilon,D_1,D_2\big)$-distortion constraints.

We say that the rate-distortion tuple $(R_1,R_2,D_1,D_2)$ is
$\epsilon$-{\em achievable} if a
$\big(2^{nR_1},2^{nR_2},n,\epsilon,D_1,D_2\big)$ distributed
code exists; for fixed parameters $\big(\epsilon,D_1,D_2\big)$,
we denote the set of all $\epsilon$-achievable pairs $(R_1,R_2)$
by $\mathcal{R}_\epsilon(D_1,D_2)$.  Then, the {\em rate region}
${\cal R}^*(D_1,D_2)$ of the two sources is defined by
\[ \mathcal{R}^*(D_1,D_2)
     \;\;\triangleq\;\; \bigcap_{\epsilon>0}\,\mathcal{R}_\epsilon(D_1,D_2).
\]

Now we are going to describe a different set of rates.  Define
$\mathbb{P}_{\mbox{\tiny LB}}$ to be the set of all probability
distributions $p(xy\hat x\hat y)$ over
$\mathcal{X}\times\mathcal{Y}\times\hat{\mathcal{X}}\times\hat{\mathcal{Y}}$,
such that:
\begin{itemize}
\item $p(xy\hat x\hat y)=p(\hat x\hat y)p(x|\hat x\hat y)p(y|\hat x\hat y)$
  (that is, $X-\hat X\hat Y-Y$ forms a Markov chain);
\item $p_{XY}=\sum_{\hat x\hat y}
  p(\hat x\hat y)p(x|\hat x\hat y)p(y|\hat x\hat y)$ ($p_{XY}$ is the source);
\item and $\expect{d_1\big(X,\hat X\big)}\leq D_1$ and
  $\expect{d_2\big(Y,\hat Y\big)}\leq D_2$.
\end{itemize}
Then, for each $p\in\mathbb{P}_{\mbox{\tiny LB}}$, define
\[ \mathcal{R}\big(D_1,D_2,p\big)\;\;\triangleq\;\;
   \left\{ (R_1,R_2)\;\left|\;\begin{array}{rcl}
           R_1 & \geq & I\big(X\wedge\hat X\hat Y\big|Y\big)[p] \\
           R_2 & \geq & I\big(Y\wedge\hat X\hat Y\big|X\big)[p] \\
           R_1+R_2 & \geq & I\big(XY\wedge\hat X\hat Y\big)[p]
           \end{array}\right.\right\},
\]
and define also $\mathcal{R}^o(D_1,D_2)\triangleq
\bigcup_{p\in\mathbb{P}_{\mbox{\tiny LB}}}\mathcal{R}\big(D_1,D_2,p\big)$.
Now we are ready to state our outer bound.

\subsection{Statement of an Outer Bound}

\medskip
\begin{center}\textcolor{gray}{\fbox{\begin{minipage}{16cm}
\vspace{-4mm}\textcolor{black}{\begin{theorem}
\label{thm:main}
\[ \mathcal{R}^*\big(D_1,D_2\big)\;\;\subseteq\;\;
   \overline{\mathcal{R}^o(D_1,D_2)}.
\]
\rend
\end{theorem}}\end{minipage}}}\end{center}\medskip

The proof of this theorem is given in Section~\ref{sec:main-proof}.
Before that, and next in Section~\ref{sec:aux-lemmas}, we develop a
number of observations and auxliary results to be used in the main
proof.

\section{Some Useful Observations and Auxiliary Results}
\label{sec:aux-lemmas}

\subsection{Distributed Rate-Distortion Codes as Constrained Source Covers}

\subsubsection{Distributed Source Covers}

An equivalent representation for a generic
$(2^{nR_1},2^{nR_2},n,\epsilon,D_1,D_2)$ code is given as follows:
\begin{itemize}
\item Two covers:
  $\mathcal{S}_1 = \big\{ \mathbf{S}_{1,i} : i=1...2^{nR_1} \big\}$
    of $\mathcal{X}^n$,
  and $\mathcal{S}_2 = \big\{ \mathbf{S}_{2,j} : j=1...2^{nR_2} \big\}$
    of $\mathcal{Y}^n$.
  Any code with encoders $f_1$ and $f_2$ can be represented in terms
  of two such covers, by considering $f_1^{-1}(i)= \mathbf{S}_{1,i}$ and
  $f_2^{-1}(j)=\mathbf{S}_{2,j}$.\footnote{Note that, strictly speaking,
  this definition is correct only when $\mathcal{S}$ is a partition.
  Occasionally we might abuse the notation and still refer to the code
  specified by a cover, with the understanding that in such cases ties
  (of the form of a source sequence being part of two different cover
  elements) are broken arbitrarily.  This should not cause any confusion.} \\
  (Note: these two covers define a cover $\mathcal{S}=\big(\mathcal{S}_1,
  \mathcal{S}_2\big)$ of $\mathcal{X}^n\times \mathcal{Y}^n$, with elements
  $\mathbf{S}_{ij} \;=\; \mathbf{S}_{1,i}\times \mathbf{S}_{2,j}$,
  for $(i,j)\in\{1...2^{nR_1}\}\times\{1...2^{nR_2}\}$.)
\item A pair of reconstruction sequences $\big(\hat{\mathbf{x}}^n(ij),
  \hat{\mathbf{y}}^n(ij)\big)=g(i,j)$ associated to each cover element
  $\mathbf{S}_{ij}$ of the product source, for all
  $(i,j)\in\{1...2^{nR_1}\}\times\{1...2^{nR_2}\}$.
\end{itemize}

In general, whenever we refer to a distributed rate-distortion code,
we use interchangeably the earlier representation in terms of two
encoders and one decoder, and this representation in terms of covers.

\subsubsection{Distributed Typical Sets}

As highlighted in the Introduction, it turns out that covers
$\mathbf{S}_{ij}$ of the product source $\mathcal{X}^n\times\mathcal{Y}^n$
are constrained beyond the requirements imposed by the fidelity
criteria.  That ``extra'' structure is described by
Proposition~\ref{prp:distributed-typicality}.

\medskip
\begin{center}\textcolor{gray}{\fbox{\begin{minipage}{16cm}
\vspace{-4mm}\textcolor{black}{\begin{proposition}
\label{prp:distributed-typicality}
For any cover $\mathcal S$ of $\mathcal{X}^n\times\mathcal{Y}^n$
defined by some $(2^{nR_1},2^{nR_2},n,\epsilon,D_1,D_2)$ distributed
rate-distortion code, and for any
$(i,j)\in\{1...2^{nR_1}\}\times\{1...2^{nR_2}\}$,
$\mathbf{x}^n\in\mathbf{S}_{1,i}$ and $\mathbf{y}^n\in\mathbf{S}_{2,j}$,
then it must be the case that 
either $(\mathbf{x}^n\mathbf{y}^n)\in\mathbf{S}_{ij}\cap
T_\epsilon^n\big(XY\big)$ or $(\mathbf{x}^n\mathbf{y}^n)\not\in
T_\epsilon^n\big(XY\big)$.  \rend
\end{proposition}}\end{minipage}}}\end{center}\medskip

{\it Proof.} This is rather straightforward.  Take any
$\mathbf{x}^n\in\mathbf{S}_{1,i}$ and $\mathbf{y}^n\in\mathbf{S}_{2,j}$.
Then:
\begin{itemize}
\item by construction,
  $\big(\mathbf{x}^n\mathbf{y}^n\big)\in\mathbf{S}_{ij}$;
\item either $\big(\mathbf{x}^n\mathbf{y}^n\big)\in
  T_\epsilon^n\big(XY\big)$ or 
  $\big(\mathbf{x}^n\mathbf{y}^n\big)\not\in
  T_\epsilon^n\big(XY\big)$ -- a tautology;
\item if $\big(\mathbf{x}^n\mathbf{y}^n\big)\in
  T_\epsilon^n\big(XY\big)$, then $\big(\mathbf{x}^n\mathbf{y}^n\big)\in
  \mathbf{S}_{ij}\cap T_\epsilon^n\big(XY\big)$, and therefore
  the proposition is proved;
\item and if instead, $\big(\mathbf{x}^n\mathbf{y}^n\big)\not\in
  T_\epsilon^n\big(XY\big)$, then the proposition is proved too.
  \tend
\end{itemize}
\medskip

Proposition~\ref{prp:distributed-typicality} formally states the
property of covers arising from distributed codes discussed informally
in the Introduction (cf.~Sec.~\ref{sec:intro-distributed-covers}): all
combinations of an $\mathbf{x}^n$ sequence in $\mathbf{S}_{1,i}$ and
a $\mathbf{y}^n$ sequence in $\mathbf{S}_{2,j}$, if they are jointly
typical, must appear in $\mathbf{S}_{ij}\cap T_\epsilon^n\big(XY\big)$
-- the decoder does not have enough information to discriminate among
such pairs.

We now introduce a new definition.
Consider any subset $\mathbf{S}\subseteq T_\epsilon^n\big(XY\big)$
for which, for any $(\mathbf{x}^n,\mathbf{y}_1^n)\in\mathbf{S}$ and
$(\mathbf{x}_1^n,\mathbf{y}^n)\in\mathbf{S}$, we have that either
$(\mathbf{x}^n\mathbf{y}^n)\in\mathbf{S}$ or
$(\mathbf{x}^n\mathbf{y}^n)\not\in T_\epsilon^n\big(XY\big)$
-- that is, the property of Prop.~\ref{prp:distributed-typicality}
holds for $\mathbf{S}$.  In this case, we say that $\mathbf{S}$ is
is a {\em distributed} typical set.

Clearly there are ``interesting'' distributed typical sets, the
concept is not vacuous:
\begin{itemize}
\item all sets of the form $\mathbf{S} = \{ (\mathbf{x}^n\mathbf{y}^n) \}$,
  with $(\mathbf{x}^n\mathbf{y}^n)\in T_\epsilon^n\big(XY\big)$,
  are distributed typical sets;
\item for any $\mathbf{S}_1\subseteq\mathcal{X}^n$ and any
  $\mathbf{S}_2\subseteq\mathcal{Y}^n$,
  $\mathbf{S}\triangleq\big[\mathbf{S}_1\!\times\!\mathbf{S}_2\big]\cap
  T_\epsilon^n\big(XY\big)$ is a distributed typical set.
\end{itemize}
The last example provides a natural way of systematically constructing
distributed typical sets.

\subsubsection{Source Covers Made of Distributed Typical Sets}

We show next that in multiterminal source coding, the source must
be covered with distributed typical sets in which each of the two
components of the set gets specified by a different encoder.

Consider a length $n$ $\big(f_1,f_2,g\big)$ code, satisfying the
$(\epsilon,D_1,D_2)$-distortion constraint of
eqn.~\eqref{eq:distortion-constraint}:
\begin{eqnarray*}
\lefteqn{P\Big( \Big\{ \big(\mathbf{x}^n\mathbf{y}^n\big) \;\Big|\;
          \big(\hat{\mathbf{x}}^n\hat{\mathbf{y}}^n\big)
          = g\big(f_1(\mathbf{x}^n),f_2(\mathbf{y}^n)\big) \,\wedge\,
          d_1\big(\mathbf{x}^n,\hat{\mathbf{x}}^n\big)
          < D_1^+ \,\wedge\,
          d_2\big(\mathbf{y}^n,\hat{\mathbf{y}}^n\big)
          < D_2^+ \Big\}
    \Big)} \\
  & \stackrel{(a)}{=} &
          P\Big( \Big\{ \big(\mathbf{x}^n\mathbf{y}^n\big) \;\Big|\;
          \big(\hat{\mathbf{x}}^n\hat{\mathbf{y}}^n\big)
          = g\big(f_1(\mathbf{x}^n),f_2(\mathbf{y}^n)\big) \,\wedge\,
          d_1\big(\mathbf{x}^n,\hat{\mathbf{x}}^n\big)
          < D_1^+ \,\wedge\,
          d_2\big(\mathbf{y}^n,\hat{\mathbf{y}}^n\big)
          < D_2^+ \Big\}
          \cap\bigcup_{(i,j)}\mathbf{S}_{ij}
    \Big) \\
  & = & P\Big(
          \bigcup_{(i,j)}
          \Big\{ \big(\mathbf{x}^n\mathbf{y}^n\big) \;\Big|\;
          \big(\hat{\mathbf{x}}^n\hat{\mathbf{y}}^n\big)
          = g\big(f_1(\mathbf{x}^n),f_2(\mathbf{y}^n)\big) \,\wedge\,
          d_1\big(\mathbf{x}^n,\hat{\mathbf{x}}^n\big)
          < D_1^+ \,\wedge\,
          d_2\big(\mathbf{y}^n,\hat{\mathbf{y}}^n\big)
          < D_2^+ \Big\}
          \cap\mathbf{S}_{ij}
    \Big) \\
  & \stackrel{(b)}{=} & P\Big(
          \bigcup_{(i,j)}
          \Big\{ \big(\mathbf{x}^n\mathbf{y}^n\big) \;\Big|\;
          d_1\big(\mathbf{x}^n,\hat{\mathbf{x}}^n(ij)\big)<D_1^+
          \,\wedge\,\mathbf{x}^n\in\mathbf{S}_{1,i}
          \,\wedge\, d_2\big(\mathbf{y}^n,\hat{\mathbf{y}}^n(ij)\big)<D_2^+
          \,\wedge\,\mathbf{y}_2^n\in\mathbf{S}_{2,j} \Big\}
    \Big) \\
  & = & P\Big( \bigcup_{(i,j)}
          \big[\mathbf{S}_{1,i}\!\times\!\mathbf{S}_{2,j}\big]
          \cap
          \big[B\big(\hat{\mathbf{x}}^n(ij),D_1^+\big)
               \!\times\!B\big(\hat{\mathbf{y}}^n(ij),D_2^+\big)\big]
         \,\Big) \\
  & \stackrel{(c)}{\geq} & 1-\dot{\epsilon},
\end{eqnarray*}
where (a) follows from 
$\big\{ \big(\mathbf{x}^n\mathbf{y}^n\big)\,\Big|\,
\big(\hat{\mathbf{x}}^n\hat{\mathbf{y}}^n\big)
= g\big(f_1(\mathbf{x}^n),f_2(\mathbf{y}^n)\big) \,\wedge\,
d_1\big(\mathbf{x}^n,\hat{\mathbf{x}}^n\big) < D_1^+ \,\wedge\,
d_2\big(\mathbf{y}^n,\hat{\mathbf{y}}^n\big) < D_2^+ \big\}
\;\subseteq\;\mathcal{X}^n\times\mathcal{Y}^n
\;\subseteq\;\bigcup_{(i,j)} \mathbf{S}_{ij}$;
(b) follows from $\mathbf{S}_{ij}=\mathbf{S}_{1,i}\times\mathbf{S}_{2,j}$;
and (c) follows from the fact
that the code under consideration satisfies the distortion constraint
of eqn.~\eqref{eq:distortion-constraint}.  We also know, from basic
properties of typical sets, that
\[ P\Big( T_\epsilon^n\big(XY\big) \Big) \;\;\geq\;\; 1-\epsilon,
\]
and so, if we define $\tilde{\mathbf{S}}_{ij}\triangleq
\big[\mathbf{S}_{1,i}\times\mathbf{S}_{2,j}\big]\cap
T_\epsilon^n\big(XY\big)$, we see that
\begin{eqnarray}
\lefteqn{P\Big( \bigcup_{(i,j)}
          \big[\mathbf{S}_{1,i}\!\times\!\mathbf{S}_{2,j}\big]
          \cap
          \big[B\big(\hat{\mathbf{x}}^n(ij),D_1^+\big)
               \!\times\!B\big(\hat{\mathbf{y}}^n(ij),D_2^+\big)\big]
          \cap
          T_\epsilon^n\big(XY\big) \,\Big)} \nonumber
\hspace{4cm} \\
  & = & P\left( \bigcup_{(i,j)}
                \tilde{\mathbf{S}}_{ij} \cap
                \big[B\big(\hat{\mathbf{x}}^n(ij),D_1^+\big)
                  \!\times\!B\big(\hat{\mathbf{y}}^n(ij),D_2^+\big)\big]
          \right) \nonumber \\
  & \geq & 1-\ddot\epsilon;
  \label{eq:distortion-constraint-2}
\end{eqnarray}
that is, since $\tilde{\mathbf{S}}_{ij}$ is a distributed typical set,
the source must be covered with the fraction of such sets contained in
pairs of balls centered at the reconstruction sequences; furthermore,
we note that each component of the distributed typical set must be
specified completely by each encoder.


\subsection{The ``Reverse'' Markov Lemma}
\label{sec:reverse-markov-lemma}

\subsubsection{The Standard Form}

Lemma~\ref{lemma:markov} is the Markov lemma as stated
in~\cite[pg.\ 202]{Berger:78}, in our own notation.

\medskip\begin{lemma}[Markov]
\label{lemma:markov}
Consider a Markov chain of the form $X-Z-Y$.  Then, for all $\epsilon>0$,
\[ \lim_{n\to\infty}
   P\Big( \big(X^n,\mathbf{y}^n\big)\in T_\epsilon^n\big(XY\big)
          \;\Big|\;
          \big(Z^n,\mathbf{y}^n\big)\in T_\epsilon^n\big(ZY\big)
    \Big) \;\;=\;\; 1,
\]
for any sequence $\mathbf{y}^n\in\mathcal{Y}^n$.
\rend
\end{lemma}\medskip

The lemma says that for {\em every} $\mathbf{y}^n\in\mathcal{Y}^n$,
{\em if} the random vector
$\big(Z^n,\mathbf{y}^n\big)\in T_\epsilon^n\big(ZY\big)$, {\em then}
the random vector $\big(X^n,\mathbf{y}^n\big)\in T_\epsilon^n\big(XY\big)$,
with high probability.  This is not true in general: if we have two pairs
of sequences $\big(\mathbf{x}^n\mathbf{z}^n\big)\in T_\epsilon^n\big(XZ\big)$
and $\big(\mathbf{z}^n\mathbf{y}^n\big)\in T_\epsilon^n\big(ZY\big)$, it
is not always the case that
$\big(\mathbf{x}^n\mathbf{z}^n\mathbf{y}^n\big)\in T_\epsilon^n\big(XZY\big)$,
and therefore that 
$\big(\mathbf{x}^n\mathbf{y}^n\big)\in T_\epsilon^n\big(XY\big)$; that
is, joint typicality is {\em not} a transitive relation.  However,
if $X-Z-Y$ forms a Markov chain, and then only in a high probability
sense, said transitivity property holds.

\subsubsection{A Converse Statement}

We are interested in a converse form of the Markov lemma.  Suppose
we are given an arbitrary distribution $p(xyz)$, whose typical
sets satisfy the constraints imposed by the Markov lemma: can we say
that $p$ itself must be a Markov chain?  It turns out the answer is
{\em almost yes} -- if some arbitrary distribution $p$ induces typical
sets like those of a Markov chain, then there must exist a Markov
chain $p'$ within $L_1$ distance $2\epsilon$ of $p$.  This statement
is made precise in the following lemma.

\medskip\begin{center}\textcolor{gray}{\fbox{\begin{minipage}{16cm}
\vspace{-4mm}\textcolor{black}{\begin{lemma}[Reverse Markov]
\label{lemma:reverse-markov}
Fix $n$, $\epsilon>0$.  Consider any distribution
$p(xyz)$ for which, for some $\mathbf{z}^n$,
\[
  T_\epsilon^n\big(X\big|\mathbf{z}^n\big)[p]
  \times T_\epsilon^n\big(Y\big|\mathbf{z}^n\big)[p]
  \;\;=\;\; T_\epsilon^n\big(XY\big|\mathbf{z}^n\big)[p].
\]
Define a Markov chain $p'(xyz)=p(z)p(x|z)p(y|z)$, with the components
$p(z)$, $p(x|z)$ and $p(y|z)$ taken from the given $p(xyz)$.  Then,
$\big|\big|p-p'\big|\big|_1\,<\,2\epsilon$.
\rend
\end{lemma}}\end{minipage}}}\end{center}\medskip

{\it Proof.}
Consider any $\mathbf{z}^n$ for which
$T_\epsilon^n\big(XY\big|\mathbf{z}^n\big)[p]\neq\emptyset$.
Since $p'$ is a Markov chain, from the direct form of the Markov
lemma we know that
\[
   T_\epsilon^n\big(X\big|\mathbf{z}^n\big)[p']
   \times T_\epsilon^n\big(Y\big|\mathbf{z}^n\big)[p']
   \;\;\subseteq_{\epsilon'}\;\;
   T_\epsilon^n\big(XY\big|\mathbf{z}^n\big)[p'];
\]
and clearly,
$\emptyset\neq
T_\epsilon^n\big(XY\big|\mathbf{z}^n\big)[p]
=
T_\epsilon^n\big(X\big|\mathbf{z}^n\big)[p]
 \times T_\epsilon^n\big(Y\big|\mathbf{z}^n\big)[p]
=
T_\epsilon^n\big(X\big|\mathbf{z}^n\big)[p']
 \times T_\epsilon^n\big(Y\big|\mathbf{z}^n\big)[p']$,
since we choose $p'$ to coincide with $p$ on the corresponding marginals,
and from our choice of $\mathbf{z}^n$.  So, this last inclusion can be
written as
\[
   T_\epsilon^n\big(X\big|\mathbf{z}^n\big)[p]
        \times T_\epsilon^n\big(Y\big|\mathbf{z}^n\big)[p]
   \;\;\subseteq_{\epsilon'}\;\;
   T_\epsilon^n\big(XY\big|\mathbf{z}^n\big)[p'],
\]
and therefore we see that
\[
   \emptyset \;\;\neq\;\;
   T_\epsilon^n\big(X\big|\mathbf{z}^n\big)[p]
        \times T_\epsilon^n\big(Y\big|\mathbf{z}^n\big)[p]
   \;\;\subseteq_{\epsilon'}\;\;
   T_\epsilon^n\big(XY\big|\mathbf{z}^n\big)[p]
    \cap T_\epsilon^n\big(XY\big|\mathbf{z}^n\big)[p'];
\]
thus, there must exist at least one triplet of sequences
$\big(\mathbf{x}^n\mathbf{y}^n\mathbf{z}^n\big)$ that
is jointly typical under both $p$ and $p'$.  So for these particular
sequences, it follows from the definition of strong typicality that
both
\[ \forall xyz: \big|\mbox{$\frac 1 n$}N\big(xyz;
   \mathbf{x}^n\mathbf{y}^n\mathbf{z}^n\big)-
   p(xyz)\big|\,<\,\mbox{$\frac\epsilon{|\mathcal{X}|
   |\mathcal{Y}||\mathcal{Z}|}$}
   \;\textrm{ and }\;
   \forall xyz: \big|\mbox{$\frac 1 n$}N\big(xyz;
   \mathbf{x}^n\mathbf{y}^n\mathbf{z}^n\big)-
   p'(xyz)\big|\,<\,\mbox{$\frac\epsilon{|\mathcal{X}|
   |\mathcal{Y}||\mathcal{Z}|}$},
\]
and therefore the $L_1$ norm of $p-p'$ can be written as
\begin{eqnarray*}
\big|\big|p'-p\big|\big|_1
  & = & \sum_{xyz}\big|p(xyz)-p'(xyz)\big| \\
  & = & \sum_{xyz}\big|p(xyz)-\mbox{$\frac 1 n$}
        N\big(xyz;\mathbf{x}^n\mathbf{y}^n\mathbf{z}^n\big)+\mbox{$\frac 1 n$}
        N\big(xyz;\mathbf{x}^n\mathbf{y}^n\mathbf{z}^n\big)-p'(xyz)\big| \\
  & \leq & \sum_{xyz}\big|\mbox{$\frac 1 n$}N\big(xyz;
        \mathbf{x}^n\mathbf{y}^n\mathbf{z}^n\big)
        -p(xyz)\big|
        +\sum_{xyz}\big|\mbox{$\frac 1 n$}
        N\big(xyz;\mathbf{x}^n\mathbf{y}^n\mathbf{z}^n\big)-p'(xyz)\big| \\
  & < & 2\epsilon,
\end{eqnarray*}
thus proving the lemma.
\tend\bigskip

Our interest in this question stems from the fact that, from the
requirement to cover a product source with distributed typical sets,
we do get constraints on the shape of various typical sets.  So we
need to characterize what distributions can give rise to those sets,
and this lemma plays an important role in that.

\subsection{Upper Bounds on the Size of Distributed Typical Cover Elements}

\medskip
\begin{center}\textcolor{gray}{\fbox{\begin{minipage}{16cm}
\vspace{-4mm}\textcolor{black}{\begin{lemma}
\label{lemma:bound-size}
Consider any $\big(2^{nR_1},2^{nR_2},n,\epsilon,D_1,D_2\big)$ distributed
rate-distortion code, represented by a cover $\mathcal{S}$.  Then, there
exists a distribution $\pi\in\mathbb{P}_{\mbox{\tiny LB}}$ such that, for
all $(i,j)\in\{1...2^{nR_1}\}\times\{1...2^{nR_2}\}$ and all $\epsilon>0$,
\[ \big|\mathbf{S}_{ij}\,\cap\,T_\epsilon^n\big(XY\big)\big|
   \;\;\leq\;\;
   2^{n(H(XY|\hat X\hat Y)[\pi]+\ddot\epsilon)},
\]
provided $n$ is large enough.  Furthermore, for all
$\mathbf{y}^n\in\mathcal{Y}^n$,
\[ \big|\mathbf{S}_{1,i}\cap T_\epsilon^n\big(X\big|\mathbf{y}^n\big)\big|
   \;\;\leq\;\;
   2^{n(H(X|\hat X\hat YY)[\pi]+\ddot\epsilon')},
\]
and similarly for all $\mathbf{x}^n\in\mathcal{X}^n$,
\[ \big|\mathbf{S}_{2,j}\cap T_\epsilon^n\big(Y\big|\mathbf{x}^n\big)\big|
   \;\;\leq\;\;
   2^{n(H(Y|\hat X\hat YX)[\pi]+\ddot\epsilon'')},
\]
also provided $n$ is large enough.
\rend
\end{lemma}}\end{minipage}}}\end{center}\medskip

{\it Proof.}  From the two-terminal rate-distortion
theorem~\cite[Thm.\ 2.2.3]{CsiszarK:81}, we know there exists a
distribution $p(xy\hat x\hat y)=p(xy)p(\hat x\hat y|xy)$, with
$p(xy)$ the given source, $\expect{d_1\big(X,\hat X\big)}\leq D_1$
and $\expect{d_2\big(Y,\hat Y\big)}\leq D_2$, and
sequences $\hat{\mathbf{x}}^n(ij)$ and $\hat{\mathbf{y}}^n(ij)$
such that, for all
$(i,j)\in\{1...2^{nR_1}\}\times\{1...2^{nR_2}\}$ and all $\epsilon>0$,
\begin{equation}
 \tilde{\mathbf{S}}_{ij}
 \;\;\subseteq\;\;
 T_\epsilon^n\big(XY\big|\hat{\mathbf{x}}^n(ij)\hat{\mathbf{y}}^n(ij)\big),
 \label{eq:const-std-rd}
\end{equation}
provided $n$ is large enough.  But since for distributed codes we
have $\tilde{\mathbf{S}}_{ij}=
\big[\mathbf{S}_{1,i}\times\mathbf{S}_{2,j}\big]\cap T_\epsilon^n\big(XY\big)$,
it follows from standard properties of typical sets that
\[ \mathbf{S}_{1,i}\cap T_\epsilon^n\big(X\big|\mathbf{S}_{2,j}\big)
   \;\;\subseteq\;\;
   T_\epsilon^n\big(X\big|\hat{\mathbf{x}}^n(ij)\hat{\mathbf{y}}^n(ij)\big)
   \mbox{\hspace{1cm}and\hspace{1cm}}
   \mathbf{S}_{2,j}\cap T_\epsilon^n\big(Y\big|\mathbf{S}_{1,i}\big)
   \;\;\subseteq\;\;
   T_\epsilon^n\big(Y\big|\hat{\mathbf{x}}^n(ij)\hat{\mathbf{y}}^n(ij)\big).
\]
Consider now a new cover $\mathcal{S}'$, having the property that
\[ \mathbf{S}'_{1,i}\cap T_\epsilon^n\big(X\big|\mathbf{S}'_{2,j}\big)
   \;\;=\;\;
   T_\epsilon^n\big(X\big|\hat{\mathbf{x}}^n(ij)\hat{\mathbf{y}}^n(ij)\big)
   \mbox{\hspace{1cm}and\hspace{1cm}}
   \mathbf{S}'_{2,j}\cap T_\epsilon^n\big(X\big|\mathbf{S}'_{1,i}\big)
   \;\;=\;\;
   T_\epsilon^n\big(Y\big|\hat{\mathbf{x}}^n(ij)\hat{\mathbf{y}}^n(ij)\big).
\]
A simple expression for the cover element $\mathbf{S}'_{1,i}$ is obtained
as follows.  Fix an index $i\in\{1...2^{nR_1}\}$:
\[\begin{array}{lrcl}
  & \forall k: \mathbf{S}'_{1,i}\cap
               T_\epsilon^n\big(X\big|\mathbf{S}'_{2,k}\big)
    & =
    & T_\epsilon^n\big(X\big|\hat{\mathbf{x}}^n(ik)\hat{\mathbf{y}}^n(ik)\big)
    \\
  \Rightarrow\hspace{6mm}
    & \bigcup_{k=1}^{2^{nR_2}}
      \mathbf{S}'_{1,i}\cap T_\epsilon^n\big(X\big|\mathbf{S}'_{2,k}\big)
    & =
    & \bigcup_{k=1}^{2^{nR_2}}
      T_\epsilon^n\big(X\big|\hat{\mathbf{x}}^n(ik)\hat{\mathbf{y}}^n(ik)\big)
    \\
  \Rightarrow
    & \mathbf{S}'_{1,i}\cap \bigcup_{k=1}^{2^{nR_2}}
          T_\epsilon^n\big(X\big|\mathbf{S}'_{2,k}\big)
    & =
    & \bigcup_{k=1}^{2^{nR_2}}
      T_\epsilon^n\big(X\big|\hat{\mathbf{x}}^n(ik)\hat{\mathbf{y}}^n(ik)\big)
    \\
  \Rightarrow
    & \mathbf{S}'_{1,i}\cap S_{\epsilon,Y}^n\big(X\big)
    & =
    & \bigcup_{k=1}^{2^{nR_2}}
      T_\epsilon^n\big(X\big|\hat{\mathbf{x}}^n(ik)\hat{\mathbf{y}}^n(ik)\big),
\end{array}\]
and since $P\big(S_{\epsilon,Y}^n\big(X\big)\big)>1-\dot\epsilon$,
$\mathbf{S}'_{1,i}$ is determined up to a set of vanishing measure;
similarly, fixing $j\in\{1...2^{nR_2}\}$, we get
$\mathbf{S}'_{2,j}\cap S_{\epsilon,X}^n\big(Y\big) = \bigcup_{l=1}^{2^{nR_1}}
T_\epsilon^n\big(Y\big|\hat{\mathbf{x}}^n(lj)\hat{\mathbf{y}}^n(lj)\big)$.

The new cover $\mathcal{S}'$ has some useful properties:
\begin{itemize}
\item for all $(i,j)$, $\mathbf{S}_{1,i}\cap S_{\epsilon,Y}^n\big(X\big)
  \subseteq\mathbf{S}'_{1,i}\cap S_{\epsilon,Y}^n\big(X\big)$ and
  $\mathbf{S}_{2,j}\cap S_{\epsilon,X}^n\big(Y\big)\subseteq
  \mathbf{S}'_{2,j}\cap S_{\epsilon,X}^n\big(Y\big)$, and therefore
  $\tilde{\mathbf{S}}_{ij}\subseteq\tilde{\mathbf{S}}'_{ij}$ as
  well, by construction;
\item for all $\big(\mathbf{x}^n\mathbf{y}^n\big)\in\tilde{\mathbf{S}}'_{ij}$,
  $d_1\big(\mathbf{x}^n,\hat{\mathbf{x}}^n(ij)\big)<D_1^+$ and
  $d_2\big(\mathbf{y}^n,\hat{\mathbf{y}}^n(ij)\big)<D_2^+$, from the
  joint typicality conditions defining $\mathbf{S}'_{1,i}$ and
  $\mathbf{S}'_{2,j}$;
\item and $P\Big(\bigcup_{ij}\tilde{\mathbf{S}}'_{ij}\Big) \geq
  P\Big(\bigcup_{ij}\tilde{\mathbf{S}}_{ij}\Big) > 1-\dot\epsilon$;
\end{itemize}
so, $\mathcal{S}'$ ``dominates'' $\mathcal{S}$ (in that every element
in $\mathcal{S}$ is contained in one element of $\mathcal{S}'$), and
$\mathcal{S}'$ satisfies the same distortion constraints that $\mathcal{S}$
does.  Therefore, an upper bound on the size of the elements in the new
cover $\mathcal{S}'$ is also an upper bound on the size of the elements
in the given cover $\mathcal{S}$.

Next we observe that new cover element $\tilde{\mathbf{S}}'_{ij}$ can be
``sandwiched'' in between two other terms:
\begin{eqnarray*}
\Big[T_\epsilon^n\big(X\big|\hat{\mathbf{x}}^n(ij)\hat{\mathbf{y}}^n(ij)\big)
     \times
     T_\epsilon^n\big(Y\big|\hat{\mathbf{x}}^n(ij)\hat{\mathbf{y}}^n(ij)\big)
     \Big]\cap T_\epsilon^n\big(XY\big)
  & \stackrel{(a)}{\subseteq} &
    \big[\mathbf{S}'_{1,i}\times\mathbf{S}'_{2,j}\big]
    \cap T_\epsilon^n\big(XY\big) \\
  & \stackrel{(b)}{\subseteq} &
    T_\epsilon^n\big(XY\big|\hat{\mathbf{x}}^n(ij)\hat{\mathbf{y}}^n(ij)\big),
\end{eqnarray*}
where (a) follows from our choice of $\mathbf{S}'_{1,i}$ and
$\mathbf{S}'_{2,j}$, and from elementary algebra of sets; and (b)
follows from eqn.~\eqref{eq:const-std-rd}, and from the product form
of distributed covers.  So, since the other inclusion always holds,
\[
   \Big[
   T_\epsilon^n\big(X\big|\hat{\mathbf{x}}^n(ij)\hat{\mathbf{y}}^n(ij)\big)
   \times
   T_\epsilon^n\big(Y\big|\hat{\mathbf{x}}^n(ij)\hat{\mathbf{y}}^n(ij)\big)
   \Big]\cap T_\epsilon^n\big(XY\big)
   \;\;=\;\;
   T_\epsilon^n\big(XY\big|\hat{\mathbf{x}}^n(ij)\hat{\mathbf{y}}^n(ij)\big)
\]
is a necessary condition on any suitable distribution $p(xy\hat x\hat y)$
whose typical sets can be used to construct the cover $\mathcal{S}'$; or
equivalently, since this must hold for every $(i,j)$,
\[ \Big[
   T_\epsilon^n\big(X\big|\hat{\mathbf{x}}^n\hat{\mathbf{y}}^n\big)
   \times
   T_\epsilon^n\big(Y\big|\hat{\mathbf{x}}^n\hat{\mathbf{y}}^n\big)
   \Big]\cap T_\epsilon^n\big(XY\big)
   \;\;=\;\;
   T_\epsilon^n\big(XY\big|\hat{\mathbf{x}}^n\hat{\mathbf{y}}^n\big),
\]
for any sequences $\hat{\mathbf{x}}^n$ and $\hat{\mathbf{y}}^n$ such
that $T_\epsilon^n\big(XY\big|\hat{\mathbf{x}}^n\hat{\mathbf{y}}^n\big)
\neq\emptyset$.  Finally we note that this last condition is equivalent
to
\begin{equation}
   T_\epsilon^n\big(X\big|\hat{\mathbf{x}}^n\hat{\mathbf{y}}^n\big)
   \times
   T_\epsilon^n\big(Y\big|\hat{\mathbf{x}}^n\hat{\mathbf{y}}^n\big)
   \;\;=\;\;
   T_\epsilon^n\big(XY\big|\hat{\mathbf{x}}^n\hat{\mathbf{y}}^n\big).
  \label{eq:const-typsets-1}
\end{equation}
This is because this last equality already forces any $\mathbf{x}^n
\in T_\epsilon^n\big(X\big|\hat{\mathbf{x}}^n\hat{\mathbf{y}}^n\big)$
and $\mathbf{y}^n\in
T_\epsilon^n\big(Y\big|\hat{\mathbf{x}}^n\hat{\mathbf{y}}^n\big)$ to
be jointly typical.  Therefore, from the reverse Markov lemma, we
conclude there exists a distribution $\pi(xy\hat x\hat y)$, which
satisfies a Markov chain of the form $X-\hat X\hat Y-Y$, such that
$\big|\big|p-\pi\big|\big|_1<2\epsilon$.

\centerline{---------------------}

Next we observe that if $\big|\big|p-\pi\big|\big|_1<2\epsilon$,
then conditionals and marginals of $p$ and of $\pi$ are also close.
Consider, for example,
$p_{\hat X\hat Y}(\hat x\hat y)=\sum_{xy}p_{XY\hat X\hat Y}(xy\hat x\hat y)$
and $\pi_{\hat X\hat Y}(\hat x\hat y)
=\sum_{xy}\pi_{XY\hat X\hat Y}(xy\hat x\hat y)$:
  \begin{eqnarray*}
  \big|\big|p_{\hat X\hat Y}(\cdot)-\pi_{\hat X\hat Y}(\cdot)\big|\big|_1
     & = & \sum_{\hat x\hat y}
           \big|p_{\hat X\hat Y}(\hat x\hat y)
                -\pi_{\hat X\hat Y}(\hat x\hat y)\big| \\
     & = & \sum_{\hat x\hat y}
           \Big|\Big(\sum_{x'y'}p_{XY\hat X\hat Y}(x'y'\hat x\hat y)\Big)
                -\Big(\sum_{x''y''}\pi_{XY\hat X\hat Y}(x''y''\hat x\hat y)
                 \Big)\Big| \\
     & = & \sum_{\hat x\hat y}
           \Big|\sum_{xy}p_{XY\hat X\hat Y}(xy\hat x\hat y)
                -\pi_{XY\hat X\hat Y}(xy\hat x\hat y)\Big| \\
     & \leq & \sum_{xy\hat x\hat y}
           \big|p_{XY\hat X\hat Y}(xy\hat x\hat y)
                -\pi_{XY\hat X\hat Y}(xy\hat x\hat y)\big| \\
     & < & 2\epsilon.
  \end{eqnarray*}
For the conditional $p_{XY|\hat X\hat Y}(xy|\hat x\hat y)$:
  \begin{eqnarray*}
  \lefteqn{\big|\big|p_{XY|\hat X\hat Y}(\cdot|\hat x\hat y)
                     -\pi_{XY|\hat X\hat Y}(\cdot|\hat x\hat y)\big|\big|_1
     \;\; = \;\; \sum_{xy} \big|p_{XY|\hat X\hat Y}(xy|\hat x\hat y)
                          -p_{XY|\hat X\hat Y}(xy|\hat x\hat y)\big|} \\
     & = & \sum_{xy} \Big|\frac{p_{XY\hat X\hat Y}(xy\hat x\hat y)}
                               {p_{\hat X\hat Y}(\hat x\hat y)}
                          -\frac{\pi_{XY\hat X\hat Y}(xy\hat x\hat y)}
                                {\pi_{\hat X\hat Y}(\hat x\hat y)}\Big| \\
     & = & \mbox{$\frac{1}{p_{\hat X\hat Y}(\hat x\hat y)
                           \pi_{\hat X\hat Y}(\hat x\hat y)}$}
           \sum_{xy} \big|p_{XY\hat X\hat Y}(xy\hat x\hat y)
                          \pi_{\hat X\hat Y}(\hat x\hat y)
                         -\pi_{XY\hat X\hat Y}(xy\hat x\hat y)
                          p_{\hat X\hat Y}(\hat x\hat y)\big| \\
     & \stackrel{(a)}{<} & \mbox{$\frac{1}{p_{\hat X\hat Y}(\hat x\hat y)
                                           \pi_{\hat X\hat Y}(\hat x\hat y)}$}
           \sum_{xy} \big|p_{XY\hat X\hat Y}(xy\hat x\hat y)
                           p_{\hat X\hat Y}(\hat x\hat y)
                          +p_{XY\hat X\hat Y}(xy\hat x\hat y)2\epsilon
                          -\pi_{XY\hat X\hat Y}(xy\hat x\hat y)
                           p_{\hat X\hat Y}(\hat x\hat y)\big| \\
     & \leq & \mbox{$\frac{1}{p_{\hat X\hat Y}(\hat x\hat y)
                              \pi_{\hat X\hat Y}(\hat x\hat y)}$}
           \sum_{xy}\Big(2\epsilon p_{XY\hat X\hat Y}(xy\hat x\hat y)
                    +p_{\hat X\hat Y}(\hat x\hat y)
                     \big|p_{XY\hat X\hat Y}(xy\hat x\hat y)
                          -\pi_{XY\hat X\hat Y}(xy\hat x\hat y)\big|\Big) \\
     & = & \mbox{$\frac{1}{p_{\hat X\hat Y}(\hat x\hat y)
                              \pi_{\hat X\hat Y}(\hat x\hat y)}$}
           \left(2\epsilon p_{\hat X\hat Y}(\hat x\hat y)
                    +p_{\hat X\hat Y}(\hat x\hat y)\sum_{xy}
                     \big|p_{XY\hat X\hat Y}(xy\hat x\hat y)
                          -\pi_{XY\hat X\hat Y}(xy\hat x\hat y)\big|\right) \\
     & \leq & \frac{4\epsilon}{\pi_{\hat X\hat Y}(\hat x\hat y)} \\
     & \triangleq & \epsilon_1,
  \end{eqnarray*}
where (a) follows from the $L_1$ bound on the marginals
$p_{\hat X\hat Y}$ and $\pi_{\hat X\hat Y}$ above; and provided both
$p_{\hat X\hat Y}(\hat x\hat y)\neq 0$ and
$\pi_{\hat X\hat Y}(\hat x\hat y)\neq 0$.  We also note that under the
assumption that
$\big|\big|p_{XY\hat X\hat Y}-\pi_{XY\hat X\hat Y}\big|\big|_1<2\epsilon$,
there exists a value $\hat\epsilon$ such that, for all
$0<\epsilon<\hat\epsilon$, it is not possible to have a pair
$(\hat x_0\hat y_0)$ such that $p_{\hat X\hat Y}(\hat x_0\hat y_0)>0$
but $\pi_{\hat X\hat Y}(\hat x_0\hat y_0)=0$, or vice versa.  This is
because $\pi_{\hat X\hat Y}(\hat x_0\hat y_0)=0$ means that for all $xy$,
$\pi_{XY\hat X\hat Y}(xy\hat x_0\hat y_0)=0$.  But if
$p_{\hat X\hat Y}(\hat x_0\hat y_0)>0$, this means there exists at
least one $x_0y_0$ such that $p_{XY\hat X\hat Y}(x_0y_0\hat x_0\hat y_0)>0$,
and as a result,
$\big|\big|p_{XY\hat X\hat Y}-\pi_{XY\hat X\hat Y}\big|\big|_1\geq
p_{XY\hat X\hat Y}(x_0y_0\hat x_0\hat y_0)$; thus, setting
$\hat\epsilon\triangleq p_{XY\hat X\hat Y}(x_0y_0\hat x_0\hat y_0)$,
we get the sought contradiction.  Thus, for all $\epsilon$ small enough,
the bound on the conditionals holds as well, and so we have
from~\cite[Thm.\ 16.3.2]{CoverT:91} that
\begin{equation}
   \Big|H\big(XY\big|\hat X=\hat x,\hat Y=\hat y\big)[p]
        -H\big(XY\big|\hat X=\hat x,\hat Y=\hat y\big)[\pi]\Big|
   \;\;<\;\;
   -\epsilon_1\log\Big(\mbox{$\frac{\mbox{\normalsize $\epsilon_1$}}{|\mathcal{X}||\mathcal{Y}|
   |\hat{\mathcal{X}}||\hat{\mathcal{Y}}|}$}\Big)
   \;\;\triangleq\;\; \epsilon_2,
  \label{eq:l1-bound-cond-entropy}
\end{equation}
and so,
\begin{eqnarray*}
\lefteqn{\Big|H\big(XY\big|\hat X\hat Y\big)[p]
              -H\big(XY\big|\hat X\hat Y\big)[\pi]\Big|} \\
  & \leq & \sum_{\hat x\hat y}
           \Big|p_{\hat X\hat Y}(\hat x\hat y)
                H\big(XY\big|\hat X=\hat x,\hat Y=\hat y\big)[p]
                -\pi_{\hat X\hat Y}(\hat x\hat y)
                H\big(XY\big|\hat X=\hat x,\hat Y=\hat y\big)[\pi]\Big| \\
  & \stackrel{(a)}{\leq} &
    \big|\hat{\mathcal{X}}\big|\cdot\big|\hat{\mathcal{Y}}\big|\cdot
    \Big|p_{\hat X\hat Y}(\hat x^*\hat y^*)
          H\big(XY\big|\hat X=\hat x^*,\hat Y=\hat y^*\big)[p]
         -\pi_{\hat X\hat Y}(\hat x^*\hat y^*)
          H\big(XY\big|\hat X=\hat x^*,\hat Y=\hat y^*\big)[\pi]\Big| \\
  & \stackrel{(b)}{\leq} &
    \big|\hat{\mathcal{X}}\big|\cdot\big|\hat{\mathcal{Y}}\big|\cdot
    \Big|\pi_{\hat X\hat Y}(\hat x^*\hat y^*)
         H\big(XY\big|\hat X=\hat x^*,\hat Y=\hat y^*\big)[p]
         +2\epsilon H\big(XY\big|\hat X=\hat x^*,\hat Y=\hat y^*\big)[p]
         \\&&\mbox{\hspace{1.7cm}}
         -\pi_{\hat X\hat Y}(\hat x^*\hat y^*)
          H\big(XY\big|\hat X=\hat x^*,\hat Y=\hat y^*\big)[\pi]\Big| \\
  & = &
    \big|\hat{\mathcal{X}}\big|\cdot\big|\hat{\mathcal{Y}}\big|\cdot
    \Big|2\epsilon H\big(XY\big|\hat X=\hat x^*,\hat Y=\hat y^*\big)[p]
    \\&&\mbox{\hspace{1.7cm}}
         +\pi_{\hat X\hat Y}(\hat x^*\hat y^*)
          \Big(H\big(XY\big|\hat X=\hat x^*,\hat Y=\hat y^*\big)[p]
               -H\big(XY\big|\hat X=\hat x^*,\hat Y=\hat y^*\big)[\pi]\Big)\Big|
         \\
  & \stackrel{(c)}{\leq} &
    \big|\hat{\mathcal{X}}\big|\cdot\big|\hat{\mathcal{Y}}\big|\cdot
    \Big(2\epsilon H\big(XY\big|\hat X=\hat x^*,\hat Y=\hat y^*\big)[p]
         +p_{\hat X\hat Y}(\hat x^*\hat y^*)\epsilon_2\Big) \\
  & \triangleq & \epsilon_3,
\end{eqnarray*}
where (a) follows from choosing $\hat x^*\hat y^*$ as the pair
$\hat x\hat y\in\hat{\mathcal{X}}\times\hat{\mathcal{Y}}$ that makes
the difference $\big|p_{\hat X\hat Y}(\hat x\hat y)
H\big(XY\big|\hat X=\hat x,\hat Y=\hat y\big)[p]
-\pi_{\hat X\hat Y}(\hat x\hat y)
H\big(XY\big|\hat X=\hat x,\hat Y=\hat y\big)[\pi]\big|$ largest;
(b) follows from
$\big|\big|p_{\hat X\hat Y}-\pi_{\hat X\hat Y}\big|\big|_1<2\epsilon$;
and (c) follows from eqn.~\eqref{eq:l1-bound-cond-entropy} above, and
from the triangle inequality.

We conclude this part of the proof by noting that completely analogous
arguments can be made to show that
\[ \Big|H\big(X\big|\hat X\hat YY\big)[p]
        -H\big(X\big|\hat X\hat YY\big)[\pi]\Big|
   \;\;\leq\;\;\epsilon_4
   \mbox{\hspace{1cm}and\hspace{1cm}}
   \Big|H\big(Y\big|\hat X\hat YX\big)[p]
        -H\big(Y\big|\hat X\hat YX\big)[\pi]\Big|
   \;\;\leq\;\;\epsilon_5.
\]

\centerline{---------------------}

We are now ready to prove our desired bounds.

Since for all $(i,j)$, $\tilde{\mathbf{S}}_{ij}
\subseteq \tilde{\mathbf{S}}'_{ij} =
T_\epsilon^n\big(XY\big|\hat{\mathbf{x}}^n(ij)\hat{\mathbf{y}}^n(ij)\big)$,
\[ \big|\tilde{\mathbf{S}}_{ij}\big|
   \;\;\leq\;\;
   2^{n(H(XY|\hat X\hat Y)[p]+\epsilon)}
   \;\;\leq\;\;
   2^{n(H(XY|\hat X\hat Y)[\pi]+\epsilon+\epsilon_3)};
\]
therefore, choosing $\ddot\epsilon\triangleq\epsilon+\epsilon_3$,
the first bound specified by the lemma follows.

For the other two bounds, fix now $\mathbf{y}^n\in\mathcal{Y}^n$.
Since $\mathcal{S}$ is a cover, there must exist at least one value
$j_0\in\{1...2^{nR_2}\}$, such that $\mathbf{y}^n\in\mathbf{S}_{2,j_0}$.
So consider any $i\in\{1...2^{nR_1}\}$, and assume $\mathbf{S}_{1,i}
\cap T_\epsilon^n\big(X\big|\mathbf{y}^n\big)\neq\emptyset$; based on
this assumption, pick any $\mathbf{x}^n\in\mathbf{S}_{1,i}\cap
T_\epsilon^n\big(X\big|\mathbf{y}^n\big)$.  This means that
$\big(\mathbf{x}^n\mathbf{y}^n\big)\in
\big[\mathbf{S}_{1,i}\times\mathbf{S}_{2,j_0}\big]\cap
T_\epsilon^n\big(XY\big)$, and therefore that
$\big(\mathbf{x}^n\mathbf{y}^n\big)\in
\big[\mathbf{S}'_{1,i}\times\mathbf{S}'_{2,j_0}\big]\cap
T_\epsilon^n\big(XY\big)$, and hence from eqn.~\eqref{eq:const-std-rd}
we have that $\big(\mathbf{x}^n\mathbf{y}^n\hat{\mathbf{x}}^n(ij_0)
\hat{\mathbf{y}}^n(ij_0)\big)\in T_\epsilon^n\big(XY\hat X\hat Y\big)$,
and therefore we conclude that
\[ \mathbf{S}_{1,i}\cap T_\epsilon^n\big(X\big|\mathbf{y}^n\big)
   \;\;\subseteq\;\;
   T_\epsilon^n\big(X\big|\hat{\mathbf{x}}^n(ij_0)\hat{\mathbf{y}}^n(ij_0)
                          \mathbf{y}^n).
\]
We also note that if $\mathbf{S}_{1,i}\cap
T_\epsilon^n\big(X\big|\mathbf{y}^n\big)=\emptyset$, then the last inclusion
holds trivially.  Thus,
\[ \big|\mathbf{S}_{1,i}\cap T_\epsilon^n\big(X\big|\mathbf{y}^n\big)\big|
   \;\;\leq\;\;
   2^{n(H(X|\hat X\hat YY)[p]+\epsilon)}
   \;\;\leq\;\;
   2^{n(H(X|\hat X\hat YY)[\pi]+\epsilon+\epsilon_4)},
\]
Therefore, choosing $\ddot\epsilon'\triangleq\epsilon+\epsilon_4$, the
second bound specified by the lemma holds.  And the third (and last)
bound follows from an argument identical to this last one.  So the lemma
is proved.
\tend\bigskip

\section{Proof of Theorem~\ref{thm:main}}
\label{sec:main-proof}

Consider any $\big(2^{nR_1},2^{nR_2},n,\epsilon,D_1,D_2\big)$ distributed
rate-distortion code, represented by a cover $\mathcal{S}$.  Then,
\begin{eqnarray*}
\lefteqn{n(R_1+R_2) \;\; \geq \;\; H\big(f_1(X^n)f_2(Y^n)\big)} \\
  & = & H\big(f_1(X^n)f_2(Y^n)\big)
        - H\big(f_1(X^n)f_2(Y^n)\big|X^nY^n\big) \\
  & = & I\big(X^nY^n\wedge f_1(X^n)f_2(Y^n)\big) \\
  & = & H\big(X^nY^n\big)
        - H\big(X^nY^n\big|f_1(X^n)f_2(Y^n)\big) \\
  & = & nH\big(XY\big)
        - \sum_{1\leq i\leq 2^{nR_1},1\leq j\leq 2^{nR_2}}
          P\big(f_1(X^n)=i,f_2(Y^n)=j\big)
          H\big(X^nY^n\big|f_1(X^n)=i,f_2(Y^n)=j\big) \\
  & \geq & nH\big(XY\big) -
        \Big[ \max_{1\leq i\leq 2^{nR_1},1\leq j\leq 2^{nR_2}}
               H\big(X^nY^n\big|f_1(X^n)=i,f_2(Y^n)=j\big)
        \Big] \\&&\mbox{\hspace{2.06cm}}
        \Big[ \sum_{1\leq i\leq 2^{nR_1},1\leq j\leq 2^{nR_2}}
               P\big(f_1(X^n)=i,f_2(Y^n)=j\big)
        \Big] \\
  & = & nH\big(XY\big)
        - \max_{1\leq i\leq 2^{nR_1},1\leq j\leq 2^{nR_2}}
          H\big(X^nY^n\big|f_1(X^n)=i,f_2(Y^n)=j\big) \\
  & \stackrel{(a)}{\geq} & nH\big(XY\big)
        - \Big[\max_{1\leq i\leq 2^{nR_1},1\leq j\leq 2^{nR_2}}
          \log\big|\tilde{\mathbf{S}}_{ij}\big|\Big]-n\epsilon_1 \\
  & \stackrel{(b)}{\geq} &
           nH\big(XY\big) - nH\big(XY\big|\hat X\hat Y\big)[\pi]
                          - n\ddot\epsilon - n\epsilon_1 \\
  & = & nI\big(XY\wedge \hat X\hat Y\big)[\pi] - n\ddot\epsilon - n\epsilon_1,
\end{eqnarray*}
where (a) follows from splitting outcomes of $X^nY^n$ into typical and
non-typical ones, and from bounding the entropy of the typical ones with
a uniform distribution; and (b) follows from Lemma~\ref{lemma:bound-size},
for some $\pi\in\mathbb{P}_{\mbox{\tiny LB}}$.

For the individual rates, we have the following chain of inequalities:
\begin{eqnarray*}
nR_1 & \geq & H\big(f_1(X^n)\big) \\
  & \geq & H\big(f_1(X^n)\big|Y^n\big) \\
  & = & H\big(f_1(X^n)\big|Y^n\big)-H\big(f_1(X^n)\big|X^nY^n\big) \\
  & = & I\big(X^n\wedge f_1(X^n)\big|Y^n\big) \\
  & = & H\big(X^n\big|Y^n\big)-H\big(X^n\big|f_1(X^n)Y^n\big) \\
  & = & nH\big(X\big|Y\big)-H\big(X^n\big|f_1(X^n)Y^n\big) \\
  & = & nH\big(X\big|Y\big)
        -\sum_{\mathbf{y}^n\in\mathcal{Y}^n}\sum_{i=1}^{2^{nR_1}}
         P\big(f_1(X^n)=i,Y^n=\mathbf{y}^n\big)
         H\big(X^n\big|f_1(X^n)=i,Y^n=\mathbf{y}^n\big) \\
  & \geq & nH\big(X\big|Y\big)
        - \Big[ \max_{i=1...2^{nR_1},\mathbf{y}^n\in\mathcal{Y}^n}
                H\big(X^n\big|f_1(X^n)=i,Y^n=\mathbf{y}^n\big) \Big]
          \\&&\mbox{\hspace{2.18cm}}
          \Big[ \sum_{\mathbf{y}^n\in\mathcal{Y}^n}\sum_{i=1}^{2^{nR_1}}
                P\big(f_1(X^n)=i,Y^n=\mathbf{y}^n\big) \Big] \\
  & = & nH\big(X\big|Y\big)
        - \max_{i=1...2^{nR_1},\mathbf{y}^n\in\mathcal{Y}^n}
          H\big(X^n\big|f_1(X^n)=i,Y^n=\mathbf{y}^n\big) \\
  & \stackrel{(a)}{\geq} & nH\big(X\big|Y\big)
        - \Big[\max_{i=1...2^{nR_1},\mathbf{y}^n\in\mathcal{Y}^n}
          \log_2\big|\mathbf{S}_{1,i}\cap
           T_\epsilon^n\big(X\big|\mathbf{y}^n\big)\big|\Big]-n\epsilon_1 \\
  & \stackrel{(b)}{\geq} & nH\big(X\big|Y\big)
           - nH\big(X\big|\hat X\hat YY\big)[\pi]
           - n\ddot\epsilon' - n\epsilon_1 \\
  & = & nI\big(X\wedge \hat X\hat Y\big|Y\big)[\pi]
           - n\ddot\epsilon' - n\epsilon_1,
\end{eqnarray*}
where (a) follows from splitting the outcomes of $X^n$ into those
that are jointly typical with the given sequence $\mathbf{y}^n$ and
those that are not, and from bounding the entropy of the typical
ones with a uniform distribution; and (b) follows from
Lemma~\ref{lemma:bound-size}.  An identical argument shows that
$nR_2\geq nI\big(Y\wedge\hat X\hat Y\big|X\big)[\pi]-n\ddot\epsilon''
-n\epsilon_1$.  And since these conditions must hold for all
$\epsilon>0$, the theorem follows.
\tend

\section{Discussion}
\label{sec:discussion}

We conclude the first part of this paper with some discussion on
the results proved so far.

\subsection{Finite Parameterization of $\mathcal{R}^o(D_1,D_2)$}

The class of distributions used to define the Berger-Tung inner bound
is given by:
\[ \mathbb{P}_{\mbox{\tiny BT}}
   \;\;\triangleq\;\;
   \left\{p_{XYUV}\left|\begin{array}{rl}
                        \bullet & p(xy)=\sum_{uv}p_{XYUV}(xyuv) \\
                        \bullet & U-X-Y-V\textrm{ is a Markov chain} \\
                        \bullet & \expect{d_1\big(X,\gamma_1(U,V)\big)}\leq D_1
                                  \textrm{ and }
                                  \expect{d_2\big(Y,\gamma_2(U,V)\big)}\leq D_2
                        \end{array}\right\}\right.,
\]
for fixed distortions $(D_1,D_2)$, source $p(xy)$, and some functions
$\gamma_1:\mathcal{U}\times\mathcal{V}\to\hat{\mathcal{X}}$ and
$\gamma_2:\mathcal{U}\times\mathcal{V}\to\hat{\mathcal{Y}}$.
To make a direct comparison
with $\mathbb{P}_{\mbox{\tiny BT}}$ easier, we rewrite
$\mathbb{P}_{\mbox{\tiny LB}}$ in terms of two
variables $U$ and $V$ as follows:
\begin{itemize}
\item Set $\mathcal{U}\triangleq\hat{\mathcal{X}}$ and
  $\mathcal{V}\triangleq\hat{\mathcal{V}}$.
\item For any $p_{XY\hat X\hat Y}\in\mathbb{P}_{\mbox{\tiny LB}}$,
  set $p_{XYUV}(xyuv)\triangleq p_{XY\hat X\hat Y}(xy\hat x\hat y)$.
\end{itemize}
Then, it is clear that $\mathbb{P}'_{\mbox{\tiny LB}}$, defined by
\[ \mathbb{P}'_{\mbox{\tiny LB}}
   \;\;\triangleq\;\;
   \left\{p_{XYUV}\left|\begin{array}{rl}
                        \bullet & p(xy)=\sum_{uv}p_{XYUV}(xyuv) \\
                        \bullet & X-UV-Y\textrm{ is a Markov chain} \\
                        \bullet & \expect{d_1\big(X,\gamma_1(U,V)\big)}\leq D_1
                                  \textrm{ and }
                                  \expect{d_2\big(Y,\gamma_2(U,V)\big)}\leq D_2
                        \end{array}\right\}\right.,
\]
again for fixed distortions $(D_1,D_2)$, source $p(xy)$, and some
functions $\gamma_1:\mathcal{U}\times\mathcal{V}\to\hat{\mathcal{X}}$
and $\gamma_2:\mathcal{U}\times\mathcal{V}\to\hat{\mathcal{Y}}$, is
just a relabeling of $\mathbb{P}_{\mbox{\tiny LB}}$.

In terms of these sets, we can state the following bounds on
$\mathcal{R}^*(D_1,D_2)$:
\begin{equation}
   \overline{\bigcup_{p\in\mathbb{P}_{\mbox{\tiny BT}}}\mathcal{R}(D_1,D_2,p)}
   \;\;\subseteq\;\;
   \mathcal{R}^*(D_1,D_2)
   \;\;\subseteq\;\;
   \overline{\bigcup_{p\in\mathbb{P}'_{\mbox{\tiny LB}}}\mathcal{R}(D_1,D_2,p)}.
  \label{eq:region-bounds}
\end{equation}
$\mathcal{R}^*(D_1,D_2)$ is not a characterization of the region of
achievable rates that we would normally consider satisfactory, in that
it is not ``computable,'' in the sense of~\cite[pg.\ 259]{CsiszarK:81}.
Yet with eqn.~\eqref{eq:region-bounds}, we have managed to ``sandwich''
the uncomputable $\mathcal{R}^*(D_1,D_2)$ region in between two
other regions, both of which are computable:
\begin{itemize}
\item in $\mathbb{P}'_{\mbox{\tiny LB}}$, $U$ and $V$ are taken
  over finite alphabets ($\mathcal{U}=\hat{\mathcal{X}}$ and
  $\mathcal{V}=\hat{\mathcal{Y}}$);
\item and in $\mathbb{P}_{\mbox{\tiny BT}}$, although we have
  not been able to find anywhere in the literature a proof that
  the cardinality of $U$ and $V$ must be finite, presumably a
  direct application of the method of Ahlswede and K\"orner should
  produce the desired bounds~\cite{AhlswedeK:75, Salehi:78}.
\end{itemize}
This is of interest because, as far as we can tell, none of the outer
bounds we have found in the literature are computable.

\subsection{Relationship to the Berger-Tung Outer Bound}

One simple sufficient condition (which unfortunately does not hold)
for proving the inclusions in eqn.~\eqref{eq:region-bounds} to be
in fact equalities would have been to show that
$\mathbb{P}'_{\mbox{\tiny LB}}\subseteq\mathbb{P}_{\mbox{\tiny BT}}$.
However, a direct comparison among these two sets is still revealing.
Consider any distribution $p$ that satisfies the constraints of both
sets (i.e., $p\in\mathbb{P}_{\mbox{\tiny LB}}\cap
\mathbb{P}_{\mbox{\tiny BT}}$), and elements $xyuv$ for which
$p(xyuv)\neq 0$.  Then, this $p$ admits two different factorizations:
\[\begin{array}{crcl}
  & p(uv)p(x|uv)p(y|uv) & = & p(xy)p(u|x)p(v|y) \\
\Leftrightarrow & p(uv)\frac{p(uv|x)p(x)}{p(uv)}\frac{p(uv|y)p(y)}{p(uv)} 
                        & = & p(xy)p(u|x)p(v|y) \\
\Leftrightarrow & p(uv|x)p(x)p(uv|y)p(y) & = & p(xy)p(u|x)p(v|y)p(uv) \\
\Leftrightarrow & p(u|x)p(v|x)p(x)p(u|y)p(v|y)p(y)
                        & = & p(xy)p(u|x)p(v|y)p(uv) \\
\Leftrightarrow & p(v|x)p(x)p(u|y)p(y) & = & p(xy)p(uv) \\
\Leftrightarrow & p(xv)p(yu) & = & p(xy)p(uv).
\end{array}\]
Clearly, any distribution in this intersection must make all
variables pairwise independent: integrate any two of them, the
other two can be expressed as the product of their marginals.

We find this observation interesting because it provides clear
evidence that our lower bound is very different in nature from the
Berger-Tung outer bound~\cite{Berger:78, Tung:PhD}.  In that bound,
the set of distributions in the outer bound (all Markov chains of
the form $U-X-Y$ and $X-Y-V$) strictly contains
$\mathbb{P}_{\mbox{\tiny BT}}$; that means, there is a subset of
the distributions in the outer bound that generates all rates we
know to be achievable.  In our bound, since
$\mathbb{P}_{\mbox{\tiny LB}}\cap\mathbb{P}_{\mbox{\tiny BT}}$
is a degenerate set, {\em none} of the distributions
in $p\in\mathbb{P}_{\mbox{\tiny LB}}$ can be used to define a code
construction based on known methods,\footnote{Except of course for
trivial cases, such as when the two sources $X$ and $Y$ are independent,
and the distortion is maximum.}
such as the ``quantize-then-bin'' strategy used in the proof of
the Berger-Tung inner bound.

\subsection{Computation of the Outer Bound}

The finite parameterization of our outer bound is an important
contribution in itself we believe, given the fact that the Berger-Tung
outer bound is not computable.\footnote{And neither is the more modern
outer bound of Wagner and Anantharam~\cite{Wagner:PhD, WagnerA:05},
also mentioned in the introduction.}  This is of interest in part
because, at least in principle, this finite parameterization renders
the problem amenable to analysis using computational methods.  Finding
an efficient algorithm for computing solutions to the optimization
problem defined by Theorem~\ref{thm:main}, similar in spirit to the
Blahut-Arimoto algorithm for the numerical evaluation of channel
capacity and rate-distortion functions~\cite{Arimoto:72, Blahut:72},
certainly is an interesting challenge in its own right.

More fundamentally though, we believe the computability of our
bound holds the key to complete a proof of the optimality of the
Berger-Tung inner bound for the problem setup of Fig.~\ref{fig:setup}:
\begin{itemize}
\item Computational methods are of interest not only because they
  lead to answers that are ``useful in practice;'' discovering
  efficient algorithms invariably requires the uncovering of structure
  in the problem.  A good example in our field: the characterization
  by Chiang and Boyd of the Lagrange duals of channel capacity and
  rate-distortion as convex geometric programs~\cite{ChiangB:04}.
\item Last but not least, an efficient algorithm to compute the
  sandwich terms in eqn.~\eqref{eq:region-bounds} provides a fallback
  strategy.  If all else fails, at least by means of numerical methods
  we can check whether, in concrete instances of the problem, the
  lower and upper bounds coincide or not.
\end{itemize}
The achievability of the set of rates defined by Theorem~\ref{thm:main},
and the effective computation of the bounds of eqn.~\eqref{eq:region-bounds},
are the main topics considered in Part II.

\bigskip\noindent{\em Acknowledgements}--In the final version.


\end{document}